\documentclass[aps,reprint,superscriptaddress,amsmath,amssymb,longbibliography]{revtex4-1}

\usepackage{graphicx}
\usepackage{amsmath}
\usepackage{bm}
\usepackage{float}
\usepackage{hyperref}

\begin{document}

\title{Information content of note transitions in the music of J. S. Bach}

\author{Suman Kulkarni}
\affiliation{Department of Physics \& Astronomy, College of Arts \& Sciences, University of Pennsylvania, Philadelphia, PA 19104, USA}
\author{Sophia U. David}
\affiliation{Department of Bioengineering, School of Engineering \& Applied Science, University of Pennsylvania, Philadelphia, PA 19104, USA}
\affiliation{Department of Psychology, Yale University, New Haven, CT 06520, USA}
\author{Christopher W. Lynn}
\affiliation{Initiative for the Theoretical Sciences, Graduate Center, City University of New York, New York, NY 10016, USA}
\affiliation{Joseph Henry Laboratories of Physics, Princeton University, Princeton, NJ 08544, USA}
\author{Dani S. Bassett}
\altaffiliation{To whom correspondence should be addressed}
\email{dsb@seas.upenn.edu}
\affiliation{Department of Physics \& Astronomy, College of Arts \& Sciences, University of Pennsylvania, Philadelphia, PA 19104, USA}
\affiliation{Department of Bioengineering, School of Engineering \& Applied Science, University of Pennsylvania, Philadelphia, PA 19104, USA}
\affiliation{Department of Electrical \& Systems Engineering, School of Engineering \& Applied Science, University of Pennsylvania, Philadelphia, PA 19104, USA}
\affiliation{Department of Neurology, Perelman School of Medicine, University of Pennsylvania, Philadelphia, PA 19104, USA}
\affiliation{Department of Psychiatry, Perelman School of Medicine, University of Pennsylvania, Philadelphia, PA 19104, USA}
\affiliation{Santa Fe Institute, Santa Fe, NM 87501, USA}

\date{\today}

\begin{abstract}

Music has a complex structure that expresses emotion and conveys information. Humans process that information through imperfect cognitive instruments that produce a gestalt, smeared version of reality. How can we quantify the information contained in a piece of music? Further, what is the information inferred by a human, and how does that relate to (and differ from) the true structure of a piece? To tackle these questions quantitatively, we present a framework to study the information conveyed in a musical piece by constructing and analyzing networks formed by notes (nodes) and their transitions (edges). Using this framework, we analyze music composed by J. S. Bach through the lens of network science, information theory and statistical physics. Regarded as one of the greatest composers in the Western music tradition, Bach’s work is highly mathematically structured and spans a wide range of compositional forms, such as fugues and choral pieces. Conceptualizing each composition as a network of note transitions, we quantify the information contained in each piece and find that different kinds of compositions can be grouped together according to their information content and network structure. Moreover, we find that the music networks communicate large amounts of information while maintaining small deviations of the inferred network from the true network, suggesting that they are structured for efficient communication of information. We probe the network structures that enable this rapid and efficient communication of information---namely, high heterogeneity and strong clustering. Taken together, our findings shed new light on the information and network properties of Bach’s compositions. More generally, our simple framework serves as a stepping stone for exploring further musical complexities, creativity and questions therein. We expect this framework to have broad applicability in understanding how information is structured in a range of complex systems.

\end{abstract}

\maketitle

\section{Introduction}
From Tibetan throat singing to Scottish piobaireachd to modern hip hop, music is a universal aspect of human culture, enjoyed by people of all ages from all around the world. It has even been proposed that music is a fundamental part of being human \cite{mithen2005the}. Though styles, sounds, and instruments vary drastically from one culture and time period to another, it is indisputable that music has had a substantial impact on the development of humans and society \cite{welch2020impact,cross2001music}. Through music we can tell stories \cite{McClary1997the}, convey messages \cite{glennie2005musical}, and imbue the strongest of emotions \cite{scherer2013music,koelsch2014,blood2001}. It is a common human experience to feel pensive or despondent after hearing a slow song in a minor key or to feel carefree or energized after hearing an upbeat song in a major key. But how does something as abstract as music communicate so much? Past literature has discussed music in terms of expectation and surprise \cite{huron2006sweet, tillmann2014, pearce2012auditory}. In order to be evolutionarily successful, our brains are adept at forming expectations based on prior events. When these expectations are contradicted by an experience, we feel surprised. With surprise can come a host of other emotions: we may feel relief when the dissonant sound we expected was actually consonant, or we may feel distress when the musical resolution we expected did not occur \cite{meyer1956emotion}. But how do we quantify these expectations and surprises? How do we mathematically formalize and measure the information conveyed by a piece of music? Fundamentally, music is comprised of fleeting and elusive sounds, and hence may appear hard to measure.

Here, we seek to extract order from music's complexity by examining music through the lens of network science. A network consists of nodes and edges---representing entities and the connections between them, respectively. Conceptualizing each note as a node and each transition between two notes as an edge, we can build a network for any piece of music \cite{ferretti2017modeling, ferretti2018complex, Liu2010, buongiorno2021musicntwrk, clynn:nat2020}. This representation enables us to use physics-based approaches to quantitatively analyze aspects of a musical piece. Using music networks, we build a framework to study the information conveyed by a piece and apply this framework to provide a comprehensive analysis of Bach’s compositions. Bach is a natural case study given his prolific career, the wide appreciation his compositions have garnered, and the influence he had over contemporaneous and subsequent composers. His diverse compositions (from chorales to fugues) for a wide range of musicians (from singers to orchestra members) often share a fundamental underlying structure of repeated---and almost mathematical---musical themes and motifs. These features of Bach's compositions make them particularly interesting to study using a mathematical framework. 

As we listen to music, we form expectations. Upon hearing a particular note, we anticipate which notes might come next based on past transitions. The less likely the outcome, the more surprised we are upon hearing it. This ``suprisal'' can be quantified by the Shannon information entropy \cite{shannon}. Ideas from information theory have led to illuminating insights in a wide range of settings, including language \cite{piantadosi2011,plotkin2000language}, social networks \cite{eckmann2004,zhao2011}, transportation patterns \cite{rosvall2005} and music \cite{cohen1962information, hiller1966information}. We draw upon these ideas to quantify the information present in the music networks. While prior research has attempted to quantitatively identify patterns and features present across different kinds of music \cite{Liu2010, gomez2014, Boon1990, liu2013}, understanding how humans perceive these patterns is more nuanced and complex than simply evaluating the structure of compositions because humans are not perfect learners. Rather, studies have consistently found that humans assimilate patterns of information presented to them through imperfect perceptual systems, resulting in slightly inaccurate representations of transition structures \cite{kahneman1982judgment, dayan1993improving, clynn:natcom2020, momennejad2017successor}. This observation raises interesting questions about the information that is perceived by a human; in particular, how does the inferred structure relate to, and differ from, the true structure of a musical piece? Further, are there any patterns in music that particularly shine through the messy process of human perception and if so, how do these patterns vary across different kinds of music? While these questions are nuanced and can depend on factors like training, recent advances in the study of how humans learn networks of information offer a valuable framework to address these questions \cite{clynn:natcom2020, howard2002distributed, gershman2012successor, clynn:pnas2020}. 

Here, we draw upon ideas from network science, information theory and cognitive science to build a framework to investigate the information conveyed by music. We then use this framework to provide a systematic analysis of music composed by J. S. Bach. We begin in Sec. \ref{sec:music_network} with a discussion of how music can be represented as a network along with details of the compositions analyzed in our work. Next, in Sec. \ref{sec:music_info}, we study the information present in the networks. We find that Bach's music networks contain more information than expected from typical (or random) transition structures. Strikingly, we also find that certain composition forms are clustered together based on their information content. We investigate how the network structure influences information content, and show that the higher information in these music networks and the differences observed across musical pieces within each compositional form can be explained by the heterogeneity in node degrees (or the number of distinct pitches that follow a given note). Next, in Sec. \ref{sec:music_perceived}, we use a maximum-entropy model for how humans perceive networks of information to examine how closely the inferred transition structure of a piece aligns with the true network structure. We hypothesize that the music networks maintain a low deviation between the inferred and true network, and this property is driven by tight clustering in the network. Additionally, we find that certain compositional forms can be distinguished based on the discrepancies between the original and the inferred network. Together, our framework introduces a fresh perspective on music, and sheds new light on properties of Bach's music. By performing a systematic study of how information in a complex system, like music, is structured and perceived by humans, our work provides insights on human creativity and how humans experience the world around them. Our study also opens up numerous interesting directions for further inquiry, which we discuss in Sec. \ref{sec:discussion}.

\section{Music as a network of note transitions}\label{sec:music_network}

\begin{figure}[h!]
\centering
\includegraphics[width = \linewidth]{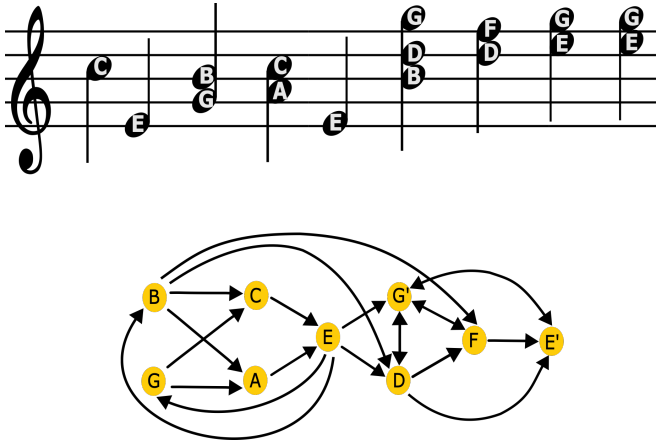}
\caption{An example of a network constructed from a musical piece using the method described in our paper. At the top, we show a toy musical piece. Below, we show the network in which notes are nodes and transitions between notes, whether isolated or played simultaneously as part of a chord, are directed edges. The direction of the edge matches the temporal direction of the transition.}
\label{fig:Fig1_NetworkExample}
\end{figure}

We note that there have been previous efforts in constructing and analyzing different network representations of music \cite{ferretti2017modeling, ferretti2018complex, Liu2010, buongiorno2021musicntwrk, clynn:nat2020}. In our study, we focus on investigating the information conveyed by note transitions in music and begin with a basic representation of the note transitions. We study a wide range of Bach's compositions including: preludes, fugues, inventions, cantatas, English suites, French suites, chorales, Brandenburg concertos, toccatas, and concertos. The audio files for these pieces were collected and read in MIDI format, from which the sequence of notes was extracted (see Methods section \ref{sec_make_nets} for further details on each compositional type and the sources for each piece). Each note present in a piece is represented as a node in the network, with notes from different octaves represented as distinct nodes. The transitions between notes are calculated separately for different instruments. If there is a transition from note $i$ to note $j$, then we draw a \emph{directed edge} from node $i$ to node $j$ (see Fig. \ref{fig:Fig1_NetworkExample}). For chords, where multiple notes occur at the same time, edges are drawn between all notes in the first chord to all notes in the second chord. To simplify our analysis, we remove any self loops in the network, thereby restricting ourselves to understanding the structure of transitions to the next \emph{different} note in the piece. We begin by examining unweighted networks of note transitions to focus on how the network structure alone impacts the information content and perception of a musical piece. After understanding the skeleton of the transitions, we then add weights to the edges based on how frequently various transitions occur. This procedure allows us to disentangle the effects of the network structure (comprising the set of possible note transitions) and edge weights (comprising the note transition probabilities). Although our emphasis has been on building a basic representation of the note transitions present in a musical piece, it is important to highlight the potential to extend this representation to capture other essential aspects of music. We expand on how future efforts could incorporate more musical realism and complexity in Sec. \ref{sec:discussion}. 

\section{Quantifying the information in networks} \label{sec:music_info}

\begin{figure}[b!]
\centering
\includegraphics[width = \linewidth]{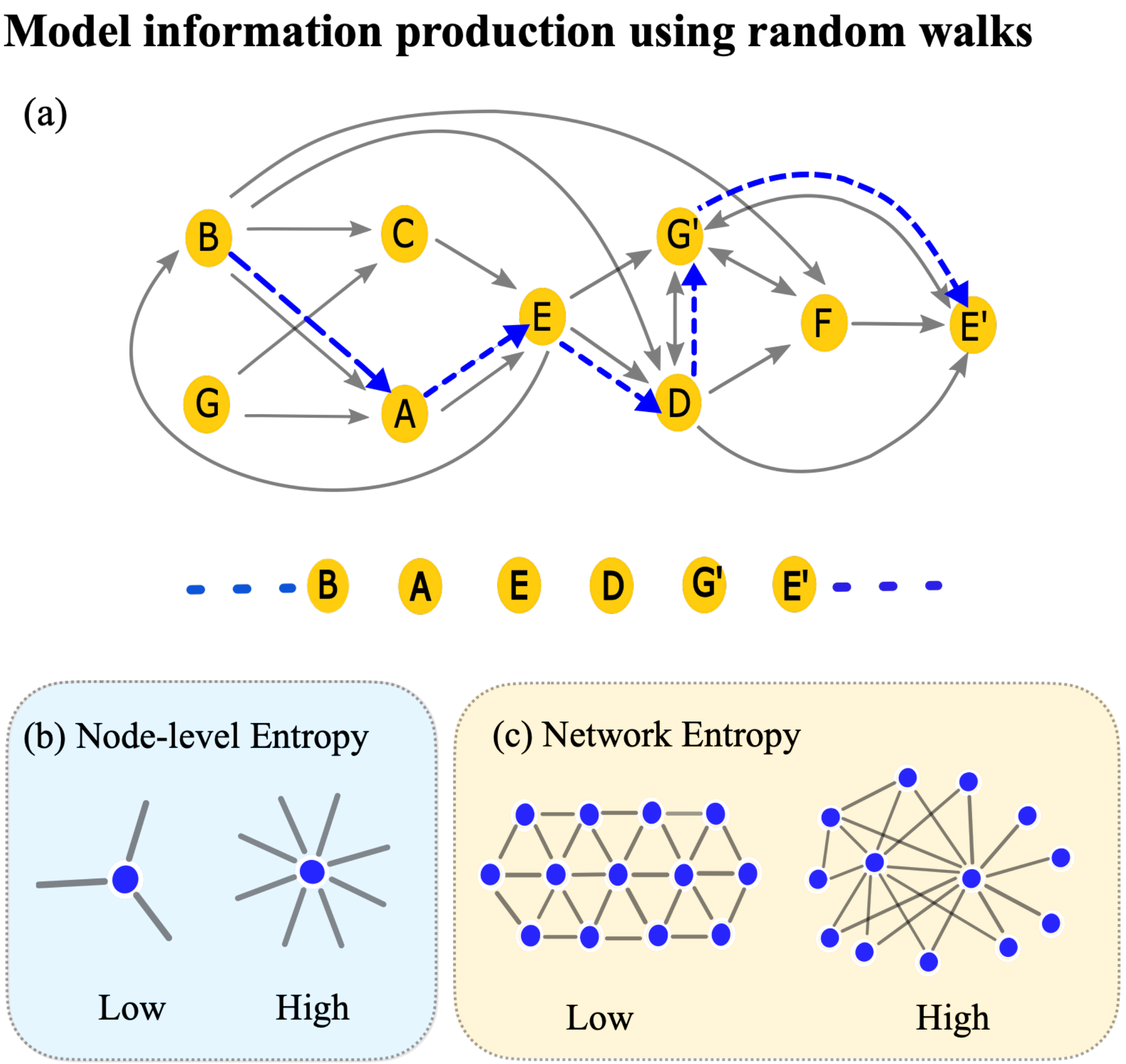}
\caption{\textbf{The model of information production using random walks.} (a) An example of a random walk on the network of note transitions is shown using the blue dotted line. At each node, the walker chooses an outgoing edge to traverse, each weighted with equal probability. This walk generates a sequence of notes as shown below. (b) The amount of information, or the \emph{entropy}, generated when a walker traverses an edge from a node depends on the degree of the node. When traversing nodes with a high versus low degree, the walker has more choices for which edge to pick and hence, such a transition generates more information. Thus, nodes with a higher degree (right) are said to have higher entropy than nodes with a low degree (left). (c) To calculate the entropy of the entire network, one needs to weigh the contribution of each node by the probability that a walker will occupy it. For networks with the same average degree, those with a wider range of degrees (right) have a higher entropy than those with a narrower range of degrees (left).}
\label{fig:Fig2_NetworkEntropy}
\end{figure}

We seek to measure the amount of information produced by a sequence of notes. Although note sequences can have long-range temporal dependencies \cite{jafari, bach_not_markov} and higher order structure \cite{lerdahl1996generative,koelsch2013processing}, as a first analytical step, we focus on the Markov transition structure. That is, we study the information contained in individual note transitions. This information is quantified by the Shannon entropy of a random walk on the network \cite{shannon, gomez2008} (Fig. \ref{fig:Fig2_NetworkEntropy}; see also the Methods section \ref{sec_entropy} for further details). Given a network of transitions, the contribution of the $i^{th}$ node to the entropy can be written in terms of the entries of the transition probability matrix $P$ as: 
\begin{equation}
    S_i = - \sum_{j} P_{ij} \; \log \, P_{ij} .
    \label{eq:node_entropy}
\end{equation}
In the case of directed unweighted networks, $P_{ij} = 1/k_i^{out}$, where $k_i^{out}$ is the out-degree of the node. Hence, for unweighted networks, the node-level entropy is $S_i = \log \, (k_i^{out})$, which is solely determined by the out-degree. 

To calculate the entropy of the entire network, the contributions of the nodes are weighted by their stationary distribution---the probability that a walker ends up at node $i$ after infinite time---which we denote by $\pi_i$ \cite{gomez2008}. The entropy of the network is then:
\begin{equation}
    S = \sum_i \pi_i S_i = -\sum_i \pi_i \sum_j P_{ij} \; \log \, P_{ij} .
    \label{eq:net_entropy}
\end{equation}
For undirected and unweighted networks, the stationary distribution has a simple analytical form $\pi_i = k_i/2E$, where $k_i$ is the degree of node $i$, and $E$ is the total number of edges. The network entropy is then:
\begin{equation}
    S = \frac{1}{2E} \sum_i k_i \log k_i .
    \label{eq:entropy_undir_unw}
\end{equation}

\begin{figure*}[t!]
  \centering
  \includegraphics[width = \linewidth]{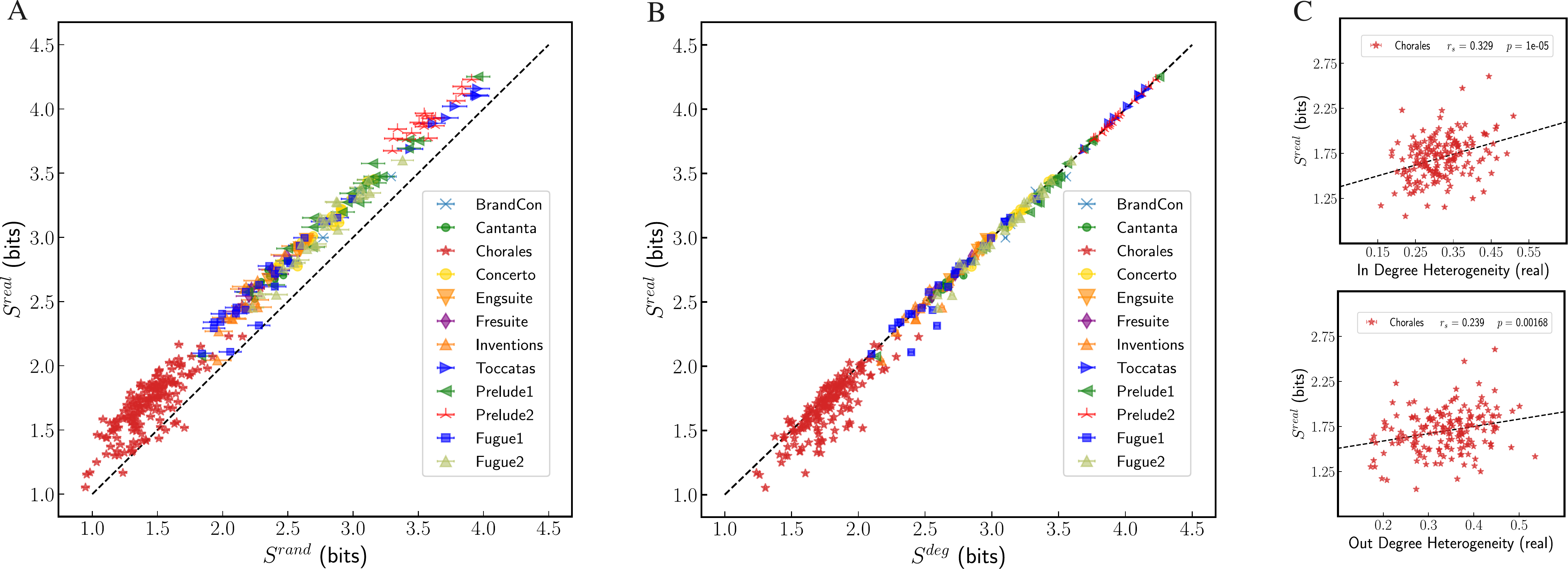}
  \caption{\textbf{Quantifying the information of Bach's music using the entropy of random walks on networks of note transitions.} (A) Entropy of Bach's music networks ($S^{\text{real}}$) compared with random networks of the same size ($S^{\text{rand}}$). We report the entropy of the corresponding random networks after averaging over 100 independent realizations. The error bars for $S^{\text{rand}}$ indicate the standard error of the sample. (B) The entropy of Bach's music networks ($S^{\text{real}}$) compared with random networks that preserve the in- and out-degree of each node ($S^{\text{deg}}$). We report the entropy of the corresponding degree-preserving random networks after averaging over 100 independent realizations. The error bars for $S^{\text{rand}}$ indicate the standard error of the sample. (C) The entropy of the chorales as a function of the average in-degree heterogeneity $H^{\text{in}} = \text{Var}(k^{\text{in}})/\langle k^{\text{in}} \rangle$ (\emph{top}) and out-degree heterogeneity $H^{\text{out}} = \text{Var}(k^{\text{out}})/\langle k^{\text{out}} \rangle$ ($\emph{bottom}$) of the networks. In panels (A) and (B), each data point represents a single piece. Color and marker indicate the type of piece, as shown in the legend. The dashed line represents the line $y=x$. In panel (C), the dotted line indicates the best linear fit, and the reported $r_s$ value is the Spearman correlation coefficient.}
  \label{fig:unweighted_entropy}
\end{figure*}
By contrast, for directed networks the stationary distribution depends on the detailed structure of the network and cannot be written in closed form. Hence, for our directed music networks, we calculate the stationary distribution numerically and use Eq. \ref{eq:net_entropy} to compute the entropy of each piece.

To understand the amount of information produced by the music networks, we compare them to randomized (or ``null'') networks of the same size; that is, networks with the same number of nodes and edges (see the Methods section \ref{sec_null_models} for details on generating null networks). This helps develop an intuition for the amount of information that networks of the same size typically contain. If the note transitions in the music networks do have distinct properties that allow them to communicate a large amount of information, then we would expect Bach's networks to contain more information than the null transition structures. By averaging over 100 random networks for each piece, we find that the real networks generally have consistently higher entropy---thereby containing more information---than their random counterparts (Fig. \ref{fig:unweighted_entropy}A). Moreover, by comparing across pieces, we observe that the different kinds of compositions cluster together based on their entropy. The chorales, typically meant to be sung by groups in ecclesiastical settings, are shorter and simpler diatonic pieces that display a markedly lower entropy than the rest of the compositions studied. By contrast, the toccatas, characterized by more complex chromatic sections that span a wider melodic range, have a much higher entropy. It is possible that the chorales' functions of meditation, adoration, and supplication are best supported by predictability and hence low entropy, whereas the entertainment functions of the toccatas and preludes are best supported by unpredictability and hence high entropy.

We know that the node-level entropy is defined only by the out-degrees of the nodes. Accordingly, it is useful to assess differences between the true networks and others wherein the node-level entropies have been fixed by preserving the true degree distribution. To perform this assessment, we compare the entropy of the real networks with another set of null models: randomized networks which preserve both the in- and out-degree of each node (see the Methods section \ref{sec_null_models} for details on generating these networks). We observe that the entropies of the networks are more or less preserved (see Fig. \ref{fig:unweighted_entropy}B). Although this preservation is expected for undirected networks (where the entropy is determined only by the degree distribution), it need not exist for directed networks (where the different stationary distributions contribute to the entropy). We therefore find that the entropy of music networks is primarily determined by their degree distributions rather than their stationary distributions. 

To gain intuition for how the entropy of note transitions depends on network structure, consider the case of unweighted and undirected networks. The network entropy takes a particularly simple form, as shown in Eq. \ref{eq:entropy_undir_unw}. Following a Taylor expansion around the average degree of the network (see the Methods section \ref{sec_entropy}), one obtains:
\begin{equation}
    S = \log \langle k \rangle + \frac{\text{Var}(k)}{2 \, \langle k \rangle ^2} + ...
    \label{eq:entropy_taylor2}
\end{equation}%
where $\langle k \rangle$ is the average degree of the network and $\text{Var}(k)$ is the variance of the degrees. To first order, we see that the entropy increases logarithmically with the average degree of the network. To second order, the entropy increases with the variance or the \emph{heterogeneity} of the degrees, such that more information will be produced by networks with heterogeneous (or broader) degree distributions. We define the degree heterogeneity as:
\begin{equation}
    H = \frac{\text{Var}(k)}{\langle k\rangle^2} .
    \label{eq:deg_het}
\end{equation}
Many networks that we encounter in our daily lives are characterized by heterogeneous degree distributions, typically with few high degree ``hub'' nodes and many low degree nodes \cite{albert2005scale, lerman2016majority, BarabasiAlbert1999scalefree}. By contrast, regular graphs---which have homogeneous degrees---produce random walks with the least entropy (see Fig. \ref{fig:Fig2_NetworkEntropy}(c)).

Where does Bach's music fall along this spectrum? We found in Fig. \ref{fig:unweighted_entropy}A that the music networks analyzed have consistently higher entropy than null networks with the same number of nodes and edges (in other words, randomized networks with the same average degree). In the Supplementary Information Sec. \ref{S1_B}, we show that this higher information content of Bach's music networks is due to higher heterogeneity in their in- and out-degree distribution; that is, the music networks are more heterogeneous in their degrees than expected from transition structures of their size, enabling them to pack more information into their structure. Since we have focused our analysis on the first-order sequential relationships among notes, which are likely common across different kinds of music, we expect this result to generalize for other kinds of music as well.

In Fig. \ref{fig:unweighted_entropy}A, we also observed that various pieces belonging to certain compositional forms were clustered together in their entropy. Consistent with this observation, we find that the pieces which are clustered together in their entropy have very similar degrees (see Supplementary Information Sec. \ref{S1_A}). Examples include English suites, French suites, and chorales. In contrast, fugues did not cluster together in their entropy as much as other composition types and displayed diverse average degrees. For the compositions that are grouped together in their entropy, we find that the differences observed among the pieces in the group can be explained by their degree heterogeneity (see Supplementary Information Sec. \ref{S1_B}). We can, for example, see this relation in the chorales where the pieces which have a higher in- and out-degree heterogeneity tend to have a higher entropy, despite having similar degrees (Fig. \ref{fig:unweighted_entropy}$C$). We note that this relationship between the entropy and degree heterogeneity holds even in our data set of \emph{directed} networks, likely because the in- and out-degrees tend to be correlated.

\section{How humans perceive networks of information}

\begin{figure}[t!]
    \includegraphics[width=\linewidth]{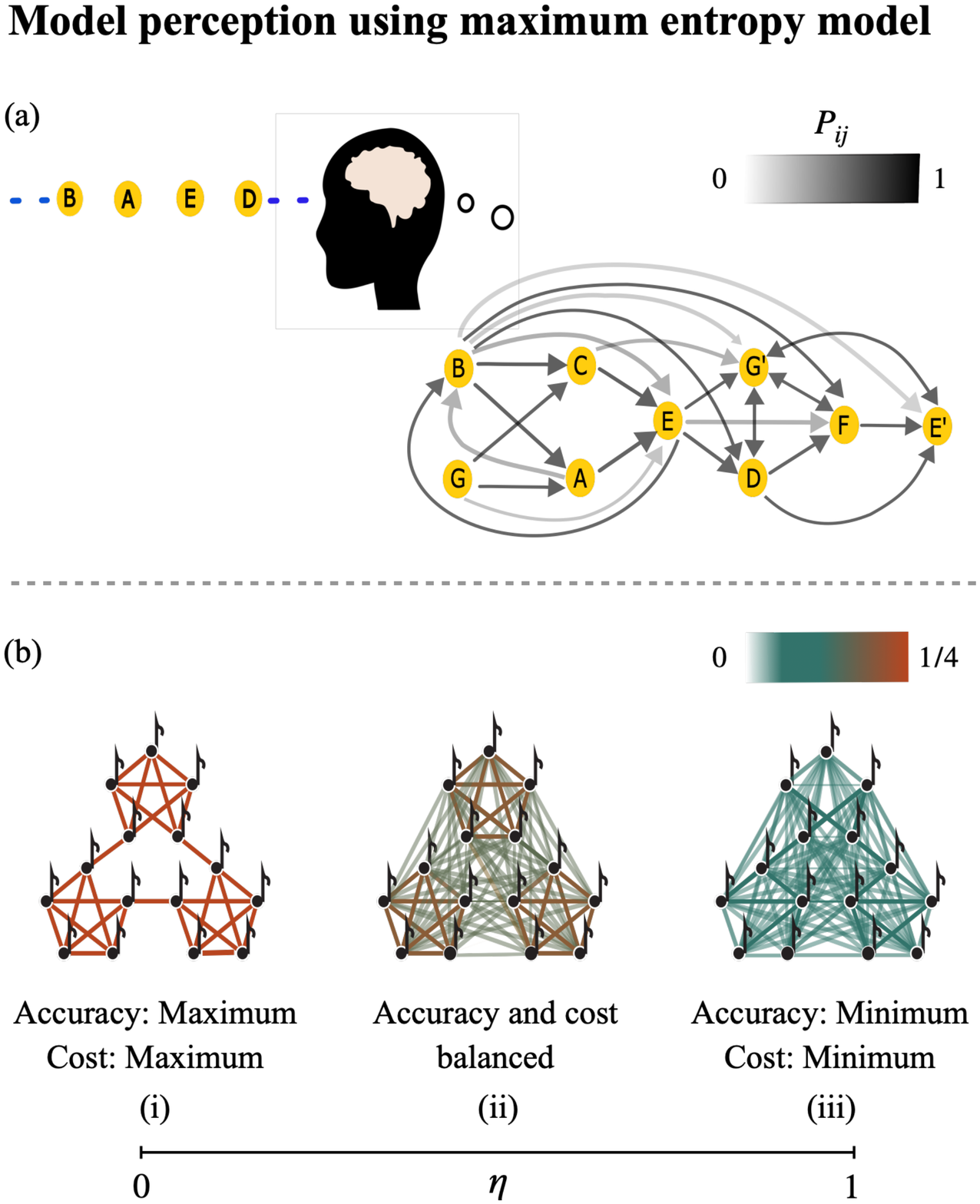}
    \caption{\textbf{How humans process networks of information.} (a) A key aspect of human communication involves receiving and assimilating information in the form of interconnected stimuli. Humans assimilate patterns of information presented to them through imperfect perceptual systems, which results in slightly inaccurate internal models of the underlying transition structure. (b) When forming internal network models of the world, humans strike a balance between accuracy and complexity. The parameter $\eta$ quantifies this trade-off between accuracy and cost. In panel (i), we see the example network built when solely maximizing the accuracy ($\eta \rightarrow 0)$, which forms a perfect representation of reality. However, building this network requires perfect memory and is computationally expensive. In panel (iii), we see the network built when solely minimizing the computational cost ($\eta \rightarrow 1)$, in which all nodes are connected to all other nodes, unlike the original network. Constructing this network does not require significant cost, but it provides no accuracy in representing the original information. Humans tend to display intermediate values of $\eta=0.80$ \cite{clynn:nat2020}, thereby constructing networks that preserve \emph{some} but not all of the true transition structure, as shown in panel (ii). Figure adapted with permission from Ref. \cite{clynn:pnas2020}.}
    \label{fig:learning_networks}
\end{figure}

A key aspect of human communication involves receiving and assimilating information in the form of interconnected stimuli---ranging from sequences of words in language and literature to melodic notes of a musical piece, and even abstract concepts. Humans assimilate this information and build representations of the underlying structure of inter-item relationships, as depicted in Fig. \ref{fig:learning_networks}a. As noted earlier, humans build these internal network models using imperfect cognitive instruments that result in slightly distorted versions of true network structures. The information that is perceived by a human is the sum of the information present in the system and the inaccuracies that stem from the imperfect cognitive processes involved in perception \cite{clynn:nat2020}. In the previous section, we focused on quantifying the actual information present in the system (see Fig. \ref{fig:Fig2_NetworkEntropy}). We will now account for the second piece: the inaccuracies that arise due to the imperfect cognitive process of perceiving information (see Fig. \ref{fig:learning_networks}).

To understand how humans learn and represent transition structures, researchers have conducted a number of experiments and introduced a range of models describing how humans internally construct transition networks \cite{clynn:natcom2020, garvert2017map, howard2002distributed, meyniel2016human, momennejad2017successor, meyniel2017brain}. A common thread across a number of these studies and models is that humans integrate transition probabilities over time, relating items that are adjacent to each other as well as those separated by transitions of length two, three, and so on \cite{newport2004learning,meyniel2016human,momennejad2017successor,clynn:nat2020}. This allows for lower computational costs and better generalizations about new information at the cost of accuracy. Here, we focus on one such model based on a free-energy principle which captures this temporal integration and inaccuracies in perception \cite{clynn:natcom2020,clynn:nat2020}. The model postulates that when constructing internal network representations of information, humans aim to  maximize the accuracy of their internal representation while simultaneously minimizing the computational cost required for its construction \cite{kahn2018, clynn:pnas2020, clynn:nat2020, clynn:natcom2020}. One the one hand, a human could learn the structure with no errors, forming a perfectly accurate network of the transitions (Fig. \ref{fig:learning_networks}b (i)) but that formation process would be computationally expensive. On the other hand, one could disregard accuracy and have the least expensive representation (Fig. \ref{fig:learning_networks}b (iii)). Most humans do something in between by recalling the sequence of transitions sometimes accurately and sometimes inaccurately, thereby forming a fuzzy perception of the true network (Fig. \ref{fig:learning_networks}b (ii)). Formally, the competition between computational complexity and accuracy can be captured by a free energy model of people's internal representation \cite{clynn:natcom2020}. The learned transition probabilities under this model ($\hat{P}$) can be written in terms of the true transition probabilities ($P$) as follows:
\begin{equation}
    \hat{P} = (1- \eta )P(I - \eta P)^{-1} ,
    \label{eq:phat}
\end{equation}
where $\eta \in [0,1] $ captures the errors in representation. A detailed derivation of this expression is provided in the Methods section \ref{sec_learning}. We emphasize the similarity of this form across multiple different theories of cognition \cite{howard2002distributed, dayan1993improving, gershman2012successor}. By relating the inferred transition structure to the true network structure, this framework enables one to explore questions about the information that a human perceives from a given network. Given our interest in such questions in the context of music, we use this model to compute the inferred network for each musical piece. We note that studies of musical expectancy have highlighted the role of statistical learning as a mechanism, alongside other factors, in musical expectancy and knowledge acquisition \cite{saffran1999,morgan2019, cheung_2020,collins2014combined}. 

For the rest of our discussion, we use the term ``inferred network" on its own to refer to the network calculated using the model of perception discussed above. Prior work indicates that, on average, humans display an $\eta = 0.80$ in large-scale online laboratory experiments \cite{clynn:nat2020}. Given a network of note transitions with transition probabilities ($P$), we use this empirically measured value to calculate the inferred network ($\hat{P}$) using Eq. \ref{eq:phat}. In the context of music, it is important to recognize that the inferred structure would naturally exhibit variations, potentially influenced by factors like an individual's level of training. Nonetheless, this framework provides interesting insights regarding the types of structures that could be considered more effective in accurately communicating information, while taking into account the limitations of human perceptual systems. We provide a discussion of how future research could expand upon our research and improve the study of information perception in music in Sec. \ref{sec:discussion}. 

\section{Quantifying discrepancies in the perception of music networks} \label{sec:music_perceived}

We are now prepared to investigate the extent to which the inferred music networks deviate from their true structure. Networks that display a low deviation between the inferred and true structure can be regarded as more effective in accurately communicating information. Hence, this framework provides insight into the communicative success of a network, from the point of view of how the network interacts with our imperfect perceptual systems. Mathematically, one can quantify the deviations between the inferred network ($\hat{P}$) and the original network ($P$) using the Kullback-Leiber (KL) divergence:%
\begin{equation}
    D_{KL}(P || \hat{P}) = - \sum_i \pi_i \sum_j P_{ij} \; \log \, \frac{\hat{P}_{ij}}{P_{ij}} ,
    \label{eq:KL_div1}
\end{equation}%
where $\pi_i$ is the stationary distribution of the original network. The lower the KL-divergence, the closer the network is to the true network, and hence the network can be considered more effective in communicating information accurately. Do Bach's musical compositions possess distinct features that result in smaller discrepancies in their perceived structure? How do pieces differ in these discrepancies? What are the structural differences between the musical pieces that lead to such differences? 

\begin{figure*}[t!]
  \centering
  \includegraphics[width = 0.8\linewidth]{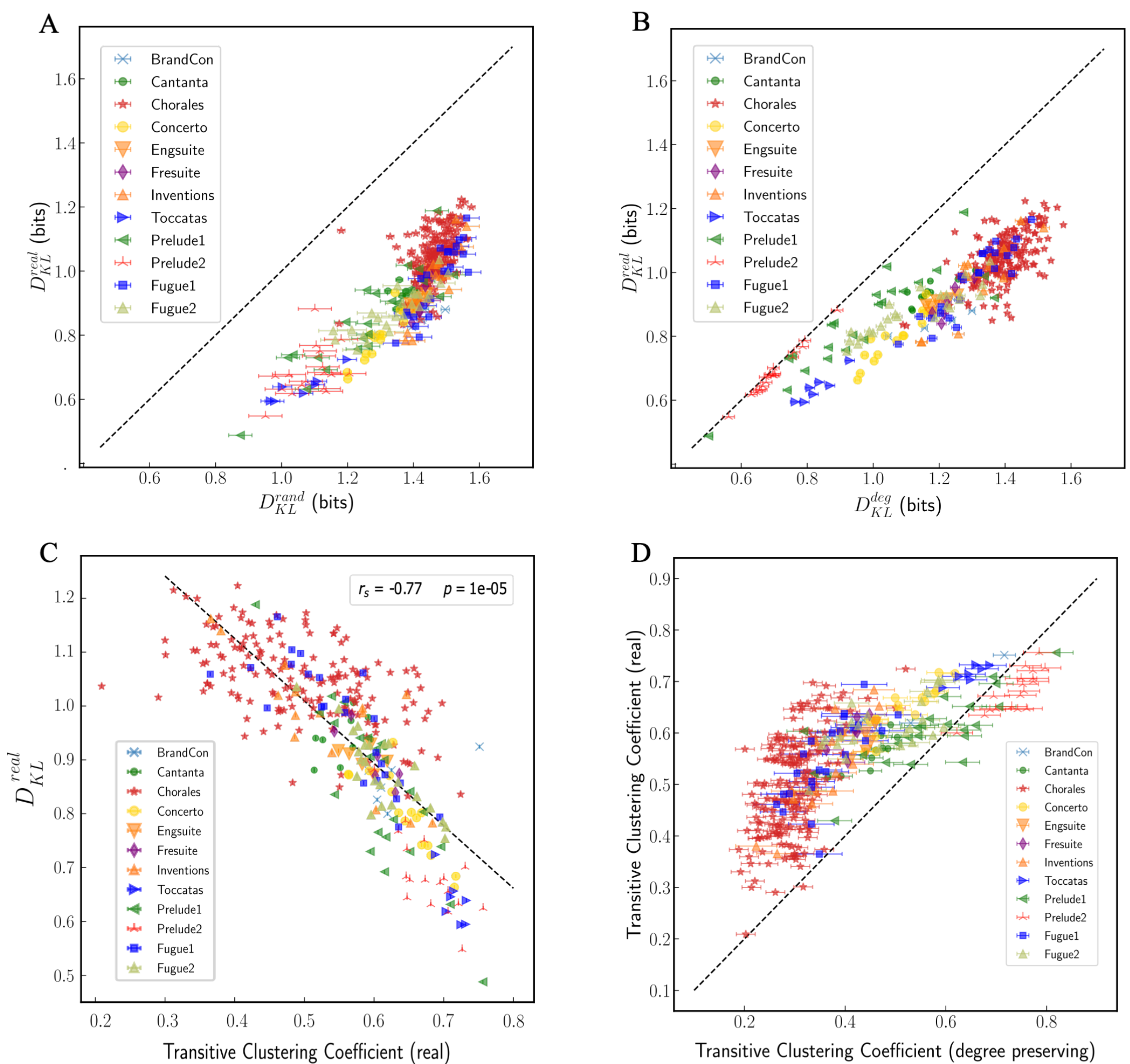}
  \caption{\textbf{Quantifying the difference between the actual information and the perceived information in Bach's music networks by calculating the KL-divergence between the actual and perceived network.} (A) KL-divergence of the real music networks ($D_{\text{KL}}^{\text{real}}$) compared with random networks of the same size ($D_{\text{KL}}^{\text{rand}}$). We report the KL-divergence of the corresponding random networks after averaging over 100 independent realizations. The error bars for $D_{\text{KL}}^{\text{rand}}$ indicate the standard error of the sample. (B) KL-divergence of the real music networks ($D_{\text{KL}}^{\text{real}}$) compared with random networks that preserve the in- and out-degree of each node ($D_{\text{KL}}^{\text{deg}}$). We report the KL-divergence of the corresponding degree-preserving random networks after averaging over 100 independent realizations. The error bars for $D_{\text{KL}}^{\text{deg}}$ indicate the standard error of the sample. (C) KL-divergence of the real music networks as a function of the transitive clustering coefficient of the network $C = \langle \Delta^{T}_i/ k_i^{\text{tot}}(k_i^{\text{tot}}-1) \rangle$. (D) The transitive clustering coefficient of the real music networks compared with random networks that preserve the in- and out-degree of each node. The dotted line indicates the line $y=x$. For the degree-preserving random networks, we report the transitive clustering coefficient after averaging over 100 independent realizations, with error bars denoting the standard error of the sample. In all the panels, each data point represents a single piece. Color and marker indicate the type of piece, as shown in the legend. The dotted line in panels (A), (B), and (D) represents the line $y=x$. }
  \label{fig:kld_figures}
\end{figure*}

To answer these questions, for each musical piece, we compute the KL-divergence between the true transition probabilities $P$ and the inferred transition probabilities $\hat{P}$. Then, to understand whether these music networks do indeed maintain low discrepancies in their inferred structure, we compare them against random networks with the same number of nodes and edges. The data confirms our intuition (Fig. \ref{fig:kld_figures}A): Bach's music networks have a lower KL-divergence than random networks of the same size. Even if we compare against null networks with the same in- and out-degree distributions, we still see that the music networks have a lower KL-divergence (Fig. \ref{fig:kld_figures}B). This finding suggests that the lower KL-divergence of these networks cannot be explained by their degree distributions alone. Additionally, we observe interesting variations in the KL-divergence among the different compositional forms (Fig. \ref{fig:kld_figures}). The chorales, at one extreme, seem to have the highest KL-divergence, while the preludes and toccatas have the lowest KL-divergence. In what follows, we attempt to identify and interpret the network properties that underlie the observed variations in the discrepancies of the inferred information across compositional forms and pieces.

\subsection{Transitive clustering coefficient}

As seen in the previous section, the discrepancies in the inferred transition structure for the music networks could not be explained by the distribution of degrees alone. For undirected networks, prior research has demonstrated that the KL-divergence between the inferred and true transition structures decreases with an increase in the density of triangles within the network \cite{clynn:nat2020}. This relationship can be demonstrated by substituting the expression for the inferred version of a network (Eq. \ref{eq:phat}) into the equation for the KL-divergence (Eq. \ref{eq:KL_div1}). We now extend this analysis to our \emph{directed} networks, with the aim of generalizing this finding. By performing this substitution, we derive the subsequent expression for the KL-divergence in terms of the original network's adjacency matrix (A):

\begin{align}
    D_{KL}(P || \hat{P}) &= - \log(1 - \eta) - \frac{\eta}{ln \; 2}\sum_i \pi_i \times \nonumber \\
    &\left\{ \sum_j A_{ij} \sum_{l} \frac{1}{k_{i}^{out}}A_{il} \frac{1}{k_{l}^{out}}  A_{lj} \right\} + \mathcal{O}(\eta^2) .
    \label{eq:KL_div2}
\end{align}
Here we see that the KL-divergence depends on a product of the form $A_{ij}A_{il}A_{lj}$, which quantifies the \emph{transitive} relationships present in the network. More explicitly, it depends on the number of directed triangles of the form $i \rightarrow j \rightarrow k$ and $i \rightarrow k$. 

To quantify the extent to which a network has clusters of this form, we introduce a measure termed the \emph{transitive clustering coefficient} of the network, defined along similar lines to the clustering coefficient of a network \cite{smallworld, Szab2004Clustering}. For each node, this quantity is measured by dividing the number of transitive triangles that node $i$ is a part of ($\Delta^{T}_{i}$) by the number of possible directed triangles:
\begin{equation}
    C_i^{T} =  \frac{\Delta^{T}_i}{\, k^{\text{tot}}_i(k^{\text{tot}}_i-1)} .
\end{equation}
Here $k^{\text{tot}}_i$ is the total degree (in + out) of the node. We average this quantity over all nodes in the network to report a single value for each piece. As indicated by Eq. \ref{eq:KL_div2}, we expect the KL-divergence of the networks to primarily be driven by the transitive clustering coefficient. This relationship is indeed evident in Fig. \ref{fig:kld_figures}$C$, where we observe that musical networks with a higher transitive clustering coefficient tend to exhibit lower KL-divergence values. In this context, we also observe that the preludes and toccatas (which demonstrated relatively lower KL-divergence values) are characterized by a larger density of transitive triangles compared to other pieces like the chorales.

A natural question that arises at this point is: What is the significance of these transitive relationships within the networks, and why do they contribute to reduced disparities between the inferred and true structure? From a cognitive science perspective, this relationship between the KL-divergence and clustering arises from the tendency of humans to count transitions of length two, as discussed previously. In a scenario where a given node $i$ is connected to node $j$ and node $j$ links to node $k$, a human learner may erroneously draw an edge between node $i$ and node $k$ in their mind. However, if the network originally had a direct link from node $i$ to node $k$, such an error would reinforce an existing edge, thereby aligning the inferred network more closely with the true network. Hence, we expect networks with high clustering to be more robust to errors made during inference. From a music perspective, interpreting these triangles is not straightforward since the networks are unweighted. Nevertheless, the presence of a large density of such triangles suggests that if there is a transition between notes $i$ and $j$, and notes $i$ and $k$, there is likely also a transition between notes $j$ and $k$. This could potentially reflect the tendency of music to form tonally stable sequences of note transitions. Substantiating these claims would require further efforts, which we elaborate on in Sec. \ref{sec:discussion}.

Analyzing the transitive clustering further, we find that the musical networks have a higher transitive clustering coefficient than degree-preserving random networks (Fig. \ref{fig:kld_figures}$D$), suggesting that this feature is not due to mere coincidence. From Fig \ref{fig:kld_figures}$D$, we make an interesting observation: the preludes appear to have a \emph{lower} transitive clustering coefficient than the corresponding null networks that preserve their size and degree distribution, while the chorale pieces generally have a higher transitive clustering coefficient than expected from null networks. We probe this further in the Supplementary Information and identify meso-scale structures that could lead to the observed differences between the compositional forms.

\section{Accounting for Note Transition Frequencies}
So far, we have focused our attention on the information content and perception of unweighted (or binary) note transition networks created from Bach's music. These networks only captured whether or not a transition exists between two notes and were not sensitive to how frequently each transition occurs. The binary networks enabled us to probe how the structure of the transitions supports effective communication. However, in many real networks, not all transitions occur with the same frequency. To reflect the different frequencies with which transitions may occur, we construct networks in which transitions are weighted according to this. For example, if note $i$ follows note $j$ 90\% of the time and note $k$ follows note $j$ 10\% of the time, the edge from node $j$ to node $i$ will be more heavily weighted than the edge from node $j$ to node $k$ (see the Methods section \ref{sec_make_nets} for further details on network construction). Adding this piece of information to the networks leads us to new questions about the role that transition weights play in communicating information to listeners. For example, how is the information generated by a random walk on the network altered by differences in the frequencies of transitions? Do these differences in frequencies reduce the discrepancies in the inferred network? 
\begin{figure*}[t!]
  \centering 
  \includegraphics[width = \linewidth]{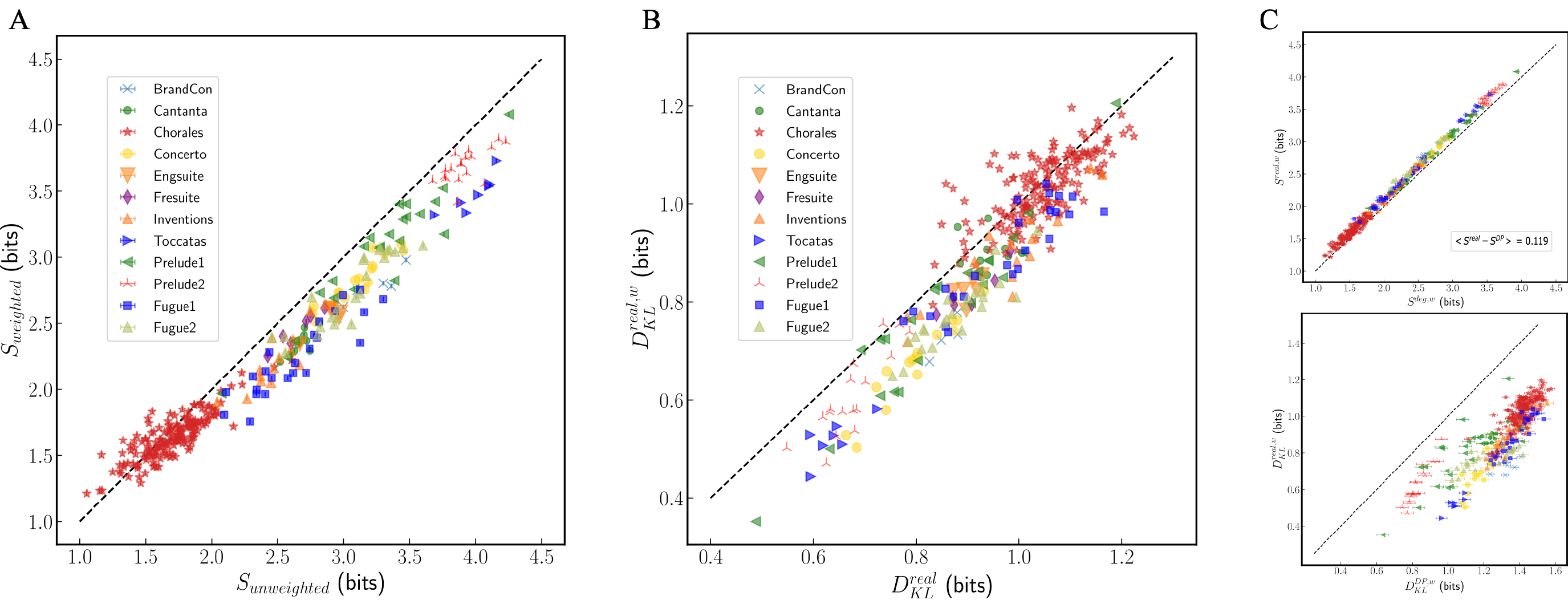}
  \caption{\textbf{Accounting for the frequencies of the note transitions in our analysis.} (A) Entropy of the weighted versions of Bach's music networks ($S_{\text{weighted}}$) compared with the corresponding unweighted versions ($S_{\text{unweighted}}$). (B) The KL-divergence of the weighted versions of Bach's music networks ($D_{KL}^{real,w}$) compared with the corresponding unweighted versions ($D_{KL}^{real}$). (C) Top: Entropy of the weighted note transition networks ($S^{\text{real,w}}$) compared with degree-preserving edge-rewired null networks ($S^{\text{deg, w}}$). Bottom: The KL-divergence of the weighted note transition networks ($D_{\text{KL}}^{\text{real,w}}$) compared with degree-preserving edge-rewired null networks ($D_{\text{KL}}^{\text{deg, w}}$). In all panels, each data point represents a single piece. Color and marker indicate the type of piece, as shown in the legend. The dashed line represents the line $y=x$. In the top figure of panel (C), we report the average deviation of the data points from the line y = x.}
  \label{fig:weighted_results}
\end{figure*}
\subsection*{Weights reduce the surprisal of transitions}
For unweighted networks, the node-level entropy of a random walk is determined solely by the out-degree $(k_{i}^{\text{out}})$, since each outgoing edge is traversed with probability $P_{ij} = 1/k_{i}^{\text{out}}$. If the edges are weighted by their transition frequencies, the $P_{ij}$'s will no longer be uniformly distributed, and each outgoing edge will not have an equal probability of being traversed. Hence, incorporating the edge weights reduces the node-level entropy. This observation is intuitive since non-uniformities in any distribution lead to decreases in entropy. However, extending this intuition to the entropy produced by the entire network is not as straightforward, since one must weigh the contribution of each node by the stationary distribution of the random walkers, which cannot be expressed in closed form for directed networks. Generally, we find that the entropy of weighted networks is still \emph{lower} than the corresponding unweighted networks (Fig. \ref{fig:weighted_results}A). This finding suggests that the different weights do indeed reduce the overall surprisal generated by the networks. 
\subsection* {Weights reduce discrepancies between the inferred network and the original network}
Incorporating the transition frequencies also helps us to understand the role that the weights play in the human inference of note transitions. We observe that the weighted networks of note transitions have \emph{lower} KL-divergence than the binary networks (Fig. \ref{fig:weighted_results}B). This observation suggests that the weights aid in forming more accurate internal representations of the transition structures, thereby reducing the discrepancies between the inferred and true structure. \\

In light of these data, we next verify the role that the network structure plays in the communicative success of weighted networks by comparing the entropy and KL-divergence of the weighted music networks with edge-rewired null networks. In the analysis on unweighted networks, we observed that the entropy was primarily driven by the degree distribution of the network and not sensitive to the precise connectivity pattern. To make this observation, we had compared the entropy of the real music networks to randomized networks that preserved the exact degree distribution of each node and hence, held the node-level entropies fixed. Along similar lines, here we make use of null models that keep the node-level entropies fixed by preserving the in- and out-degree of each node \emph{and} the out-weights at each node (see the Methods section for details on the null models). By comparing the entropy of the weighted music networks to the degree-preserving weighted null models, we see that the entropies of real networks are still more or less unchanged, although the real networks have marginally higher entropies than the null networks (Fig. \ref{fig:weighted_results}$C$, top). These results support our conclusion that the entropy in the real networks is still primarily driven by their degree distribution. When we compare the KL-divergence of the real weighted networks with the degree-preserving weighted null models, we find that the real networks have a lower KL-divergence than the corresponding null networks (Fig. \ref{fig:weighted_results}$C$, bottom). Together, these results suggest that incorporating the weights into our network analysis does not alter our results on the effects of network structure qualitatively. 

Accounting for the note transition frequencies in our network model leads to several interesting lines of inquiry. For instance, is it the specific distribution of weights that improves the accuracy of the inferred music networks? Future work could evaluate this possibility by comparing the KL-divergence of the weighted networks with a class of null models that preserve the skeleton of the network, but permute the edge weights. It would also be interesting to test whether higher edge weights are concentrated in triangular clusters of the network, offering a potential explanation for the lower KL-divergence of the weighted networks compared to the binary networks. 

\section{Conclusions and future directions} \label{sec:discussion}

Across language, literature, music and even abstract concepts, humans demonstrate the remarkable ability to identify patterns and relationships from sequences of items---an essential aspect of information sharing and communication \cite{saffran1996statistical, fiser2002statistical, romberg2010statistical, clynn:natcom2020, clynn:pnas2020}. Here, we draw upon ideas from network science, information theory and statistical physics to build a framework that serves as a stepping stone for studying the information conveyed by a musical piece. We use this framework to analyze networks of note transitions in a wide range of music composed by J. S. Bach. For each musical piece, we construct a network of note transitions by drawing directed edges between notes that are played consecutively. We then quantify the amount of information generated by the network structure and find that different compositional forms can be grouped together based on their information entropy. We relate the information content of each piece to its network structure, enabling us to gain insight into the structural properties of various pieces. Next, inspired by recent progress in the field of statistical learning which demonstrates how humans infer transition structures across visual and auditory domains \cite{clynn:natcom2020, fiser2002statistical, morgan2019, saffran1999}, we use a computational model \cite{clynn:nat2020, clynn:natcom2020} for how humans learn networks of information to compute the average ``inferred" network structure for each piece. We then quantify the discrepancies between the inferred and true transition structures under this model. Here too, we observe interesting differences among the pieces, which we attribute to differences in the clustering of the networks. Finally, we study how the frequencies of transitions influence the information content and perception of the musical pieces, by weighing the transitions by the number of times they occur. We find that the weights reduce the overall entropy or surprisal of the transitions, and also reduce the deviations between the inferred and actual network, suggesting that the weights aid in accurate inference of these transition structures.

\begin{figure}
  \centering
  \includegraphics[width=\linewidth]{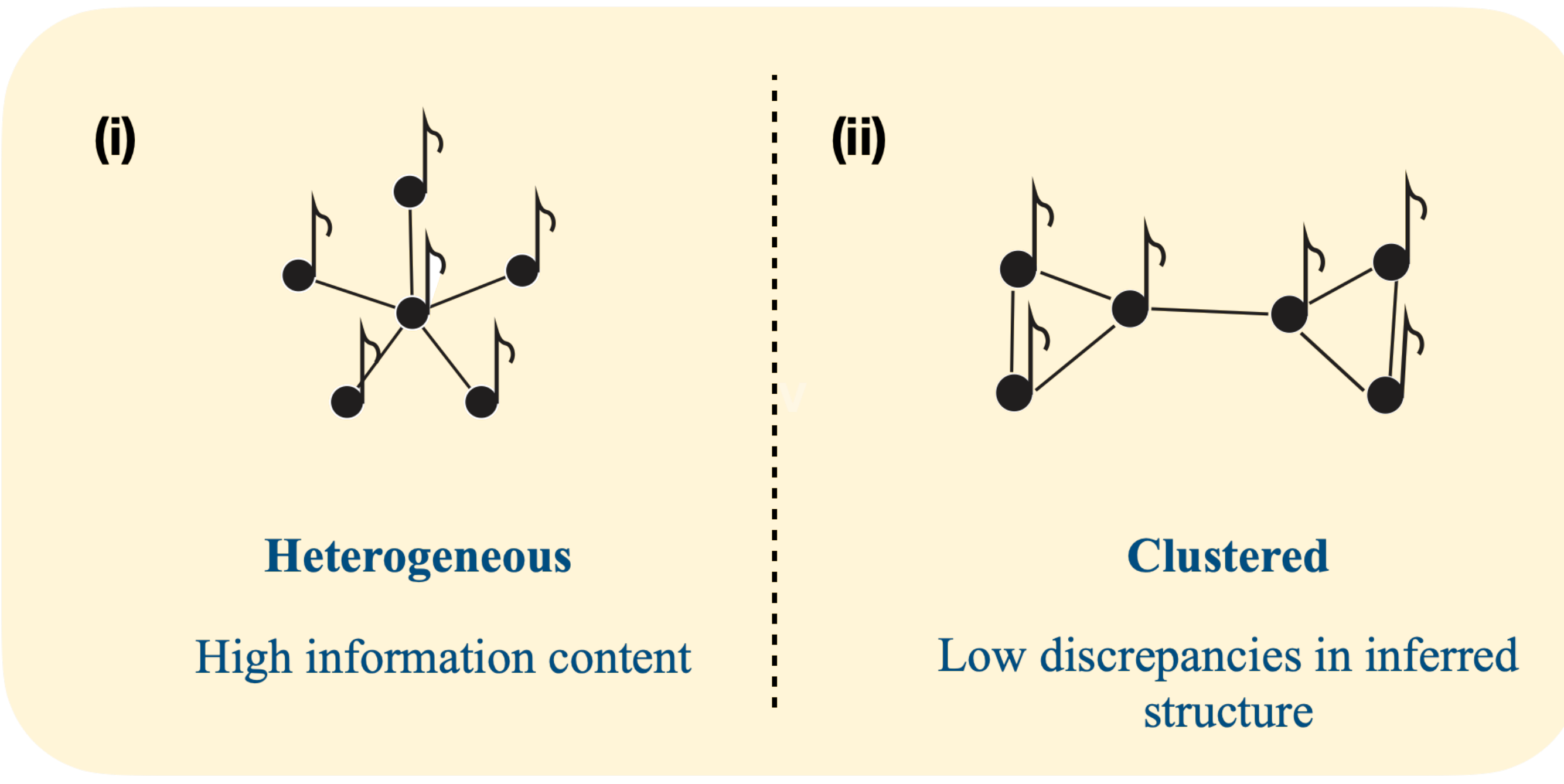}
  \caption{\textbf{Network structures that support effective communication of information.} Networks with a larger variance or heterogeneity in their node degrees, as shown in the top panel, pack more information into their structure and have a higher entropy. Clustering in the network, as shown in the bottom panel, makes the structure more resilient to errors made by humans when building an internal representation of the information, allowing the network to be inferred more accurately. Together, these structures convey a large amount of information that can be learned by humans more accurately, and are hence more efficient for communication.}
  \label{fig:efficient_networks}
\end{figure}

Furthermore, we find that the music networks contain more information and maintain lower discrepancies in the inferred structure than expected from typical transition structures of the same size. This provides us insight into features that make networks of information effective at communicating information. In general, networks which are denser (have a higher average degree) produce more information (have a higher information entropy). For networks of comparable average degree, more heterogeneous (higher variance in degree distribution) structures produce more information than those that are more regular or homogeneous in their degree (Fig. \ref{fig:efficient_networks}(i)). Moreover, networks which have a high degree of clustering maintain a lower divergence from human expectations (Fig. \ref{fig:efficient_networks}(ii)). Together, these findings suggest that for networks of a given size, rapid and accurate communication of information is supported by structures that are simultaneously heterogeneous and clustered (Fig. \ref{fig:efficient_networks}). Notably, such structures are widely prevalent across complex systems \cite{albert2005scale, lerman2016majority, BarabasiAlbert1999scalefree, watts1998collective, cancho2001small}.

We hope that our framework inspires further exchange between physics, cognitive science, and musicology. On a broader scale, our study also adds to investigations on how information in complex systems is structured. To conclude, we highlight a number of exciting directions for future inquiry and outline ways in which our framework can be expanded upon and improved.

\subsection*{Future directions}

A natural follow-up to this analysis would be to examine works of other composers---particularly works outside the Western tradition. This also prompts questions aimed at assessing how various styles or genres of music differ \cite{zivic2013, carlos2009, simonton1984}. In particular, what are the key features by which a listener distinguishes between music from two eras, say the Classical and the Romantic eras? How do the differences in structure then impact how the piece is perceived by a listener? Consequentially, a quantitative assessment of musical compositions like ours raises the intriguing possibility of identifying works of a composer or genres that may not be \emph{a priori} obvious to musicologists. 

Systematically analyzing the information that we extract from complex systems also provides us with new tools to understand human creativity and experiences. A question that often arises in the context of how humans experience music is: What makes a musical composition appealing to the human ear? While individual preferences in music can vary widely and is highly subjectively, there is still a general agreement on certain composers being considered ``influential'' or ``great''. This fact raises the possibility that there may be some inherent qualities that are common to musical pieces which are widely considered appealing. Identifying such features might give us insight into the creative process of composing music and also complement existing work using AI to generate music \cite{thickstun2016learning, liu2014bach}. Several attempts have been made to identify such patterns. For example, Ref.\,\cite{Liu2010} analyzed note transition networks in certain compositions by Bach, Chopin, and Mozart as well as Chinese pop music, and suggested that ``good'' music is characterized by the small-world property \cite{smallworld} and heavy-tailed degree distributions. On the other hand, Ref.\,\cite{gomez2014} studied selected compositions from Bach's Well-Tempered Clavier and found non-heavy-tailed degree distributions, suggesting that such distributions are not necessary for music to be appealing. It would be interesting to devise future experiments to determine whether our findings relate to the aesthetic or emotional appeal of a piece. In our study, we found that Bach's music networks had a higher number of transitive triangular clusters, enabling them to be learned more efficiently than arbitrary transition structures. Are pieces with a larger number of these triangles also more appealing to a listener? Future work assess this possibility by conducting experiments that ask people to rate Bach's compositions and analyzing whether these ratings correlate with the presence of triangular clusters. More generally, our work focuses not solely on the information inherent in the transition structure of music, but also on how the information in this transition structure is perceived by a human listener. This framework might be useful in studying cognitive aspects of music and in bridging patterns observed in data with cognitive theories of music.

In future work, it also would be interesting to extend our analysis to examine how music networks evolve with time. There are three potentially interesting lines of inquiry here: First, how do the entropy and KL-divergence of a musical piece change as the piece progresses? Does this temporal change differ among the various compositional forms? Second, how has the music of a specific composer (whether Bach or otherwise) changed over the course of their lifetime? Has it become more intricate and complex, holding more information? Perhaps as the composer gains experience, their compositions convey information more efficiently and accurately, as reflected in a reduced KL-divergence? If the exact dates of when each piece was composed were known, then the framework used in our paper might provide answers to these questions. Third, how has music of a given genre, say classical music, changed over the years across composers? Ref.\,\cite{liu2013}, for example, studied the fluctuation in pitch between adjacent notes in compositions by Bach, Mozart, Beethoven, Mendelsohn, and Chopin, and found that the largest pitch fluctuations of a composer gradually increased over time from Bach to Chopin. As mentioned earlier, it would be interesting to expand our analysis to different composers, and see how the information and expectations vary across composers and time. 

Lastly, we also identify limitations within our analysis that highlight directions for further effort. First, our work relies on a simplistic representation of music that could be expanded to incorporate more musical realism and complexity. For instance, one could account for differences in timbre, the intervals between notes, or even fused notes or chords, which are known to play a key role in music perception \cite{dowling1981importance, parncutt1989psychoacoustics}. Second, while we have focused on the information present in first-order sequential relationships among the notes, future work could capture higher-order correlations, hierarchies, and more intricate structures inherent in music \cite{jafari, bach_not_markov,lerdahl1996generative,koelsch2013processing}. Recent advances in studying higher-order dependencies and structures present in networks offer a promising approach to capturing this complexity \cite{xu2016representing, lambiotte2019networks, yin2018higher}. Incorporating such subtleties would not only improve our understanding of how the networks are structured, but also how they are perceived. Expanding on this understanding, it would be beneficial to conduct targeted experiments that specifically address and build models of the perception of distinct musical attributes. Further, exploring the variability of music perception among individuals, considering factors such as musical training or cultural influences would also be interesting. 

The aforementioned and ensuing directions would expand our capacity to address more specific questions regarding the composer idiosyncrasies, era characteristics, and genre discussed earlier. As such, our work offers a flexible framework that can be utilized by a wide range of scholars both in and outside of physics. Beyond music, our study can also be extended to a range of complex systems present around us---such as language and social networks. For example, one could analyze works of literature and ask: Does the entropy of noun transitions in various works of Shakespeare differ based on their genre? More specifically, does the information content and learnability of noun transitions or relationships between characters differ between tragedies and comedies? By providing an example of a systematic and comprehensive analysis of the actual and perceived information in music, our study complements and adds to the rich study of language, music, and art as complex systems \cite{stiller2003small, choi2007directed, gomez2014}.

\section*{Acknowledgments}
We thank Chris Macklin for an early conversation on this topic and audience members who have asked probing questions about our earlier work in communication networks. These interactions motivated our continued investigation in this space. We would also like to thank Sandip Varkey, Shubhankar Patankar and Aaron Winn for useful conversations and feedback on earlier versions of this manuscript. This particular research was primarily supported by the Army Research Office award number DCIST-W911NF-17-2-0181 and the National Institutes of Mental Health award number 1-R21-MH-124121-01. D.S.B. would also like to acknowledge additional support from the John D. and Catherine T. MacArthur Foundation, the Alfred P. Sloan Foundation, the Institute for Scientific Interchange Foundation, and the Army Research Office (Grafton-W911NF-16-1-0474). The content is solely the responsibility of the authors and does not necessarily represent the official views of any of the funding agencies.

\appendix
\section{Further details on data and methods}

\subsection{Data Collection and Network Construction}\label{sec_make_nets}
The music files were collected in the MIDI format from various sources. The sources for the compositions analyzed are as follows: preludes \cite{bc, ks}, fugues \cite{bc, ks}, inventions\cite{bc, ks}, cantatas\cite{bcw}, English suites\cite{sm}, French suites\cite{sm}, four-part chorales\cite{ks}, Brandenburg concertos\cite{ks}, toccatas\cite{sm}, and concertos\cite{sm}. The preludes and fugues are split based on whether they belong to the first or second part of The Well-Tempered Clavier, and are labelled `1' or `2'. Certain compositions consist of different movements and our data set has separate MIDI files for each movement. We analyze each movement separately and average our measurements over them to yield a single measured quantity for each piece, as indexed by a unique BWV number. In the case of the chorales, we analyzed the 186 four-part chorales in BGA Vol. 39 with BWV number 253-438.

The MIDI files were read in MATLAB using the \texttt{readmidi} function in MATLAB \cite{midi} to obtain information about the notes being played. Different instruments in a piece are stored in separate channels within each data file. The transitions between notes are calculated separately for each instrument or track. We assign each note present in a piece a node in the network, and notes from different octaves are assigned distinct nodes. We then draw an edge from note $i$ to note $j$ if there is a transition between them. If there are multiple notes being played at a single time $t$ (as is the case with chords), edges are drawn from the previously played note to \emph{all} notes at time $t$, and from all the notes being played at time $t$ to the subsequent note(s). This procedure gives us a directed binary network of note transitions. The code and data used to construct the networks is available at \cite{SK_code}. We also construct weighted versions of these networks, where each edge is weighted by the number of times the corresponding transition occurs.  

\subsection{Entropy of random walks on networks}\label{sec_entropy}

We use random walks to model how a sequence of information is generated from an underlying network of information. Under this model, a walker traverses the network by picking an outgoing edge to traverse at each node. Given a network with adjacency matrix $A$ and matrix element $A_{ij}$, the probability that a walker transitions from node $i$ to node $j$ in a standard Markov random walk is $P_{ij} = A_{ij}/k_i^{out}$, where $k_i^{out} = \sum_{j} G_{ij}$ is the out-degree of a node. We are interested in quantifying how much information is contained in the resulting sequence, which is captured by the entropy of the random walk:
\begin{equation*}
    S = - \sum_i \pi_i \sum_j P_{ij} \, \log \, P_{ij} ,
\end{equation*}
where $\pi$ is the stationary distribution of the walkers, which satisfies the condition $P \pi = \pi$. For the simplest possible case of an undirected and unweighted network, $P_{ij} = 1/k_i$ and $\pi_i = k_i/2E$, where $k_i$ is the degree of the $i^{th}$ node and $E = \sum_{i,j} A_{ij}/2$ is the total number of edges. The entropy in this case simplifies to:
\begin{equation}
   S = \frac{1}{2E} \sum_{i} k_{i} \; \log \; k_i = \frac{\langle k \; \log \; k \rangle}{\langle k \rangle}  .
\end{equation}
We can apply a Taylor expansion to this expression around the average degree of the network, and thereby obtain:
\begin{equation}
    S = \log \langle k \rangle + \frac{\text{Var(k)}}{2 \, \langle k \rangle ^2} + ...
    \label{eq:entropy_taylor_expansion}
\end{equation}
Hence we find that the entropy of random walks increase logarithmically with the average degree of the network. Additionally, it grows as the variance of the degrees increases. This formalization enables us to relate the information content of various music networks to their network structure. The code used to measure the entropy of random walks on the networks analyzed is available at \cite{SK_code}.

\subsection{Model for how humans learn networks}\label{sec_learning}
As discussed in the main text,  humans do not infer the transition probabilities of sequences of information with perfect accuracy due to imperfections in their cognitive processes. Studies have consistently found that in forming internal representations of transition structures, humans integrate transition probabilities over time \cite{newport2004learning,meyniel2016human,momennejad2017successor,clynn:nat2020}. This process results in humans connecting items in the sequence that are not directly adjacent to each other. Mathematically, we can express the inferred transition structure $\hat{P}$ in terms of the true transition structure $P$ under this model of fuzzy temporal integration as:
\begin{equation}
    \hat{P} = \sum_{\Delta t=0}^{\infty} f(\Delta t) P^{\Delta t+1} ,
    \label{eq:phat_1}
\end{equation}
where $f(\Delta t)$ is the weight given to the higher powers of $P$ and is a decreasing function of $\Delta t$ such that longer-distance associations contribute less to a person’s network representation.
The functional form of $f(\Delta t)$ is obtained using a free energy model described in Ref. \cite{clynn:natcom2020}. This model suggests that when forming internal representations of information, each human arbitrates a trade-off between accuracy and cost. The optimal distribution for $f(\Delta t)$ under this model is then a Boltzmann distribution with a parameter $\beta$ that quantifies the trade-off between cost and accuracy in forming an internal representation of the information:
\begin{equation}
    f(\Delta t) = e^{-\beta \Delta t}/Z ,
\end{equation}
where $Z = \sum e^{-\beta \Delta t} = (1- e^{-\beta})^{-1}$ is a normalization constant. Substituting this expression to simplify Eq. \ref{eq:phat_1}, we obtain an equation that relates the inferred transition probabilities $\hat{P}$ to the true transition probabilities $P$:
\begin{align}
    \hat{P} =& (1- e^{-\beta})^{-1} \sum_{\Delta t=0}^{\infty} e^{-\beta \Delta t} P^{\Delta t+1} \nonumber \\
    =& (1- \eta )P(I - \eta P)^{-1} ,
    \label{eq:phat_2}
\end{align}
where $\eta = e^{-\beta}$. Prior work has estimated the value of $\eta$ to be 0.8 from large-scale online experiments in humans \cite{clynn:nat2020}. Using this measured value of $\eta$, we use Eq. \ref{eq:phat_2} to calculate the inferred network for any given music network (code available at \cite{SK_code}).

\subsection{KL-divergence}
To quantify how much the distorted learned transition structure $\hat{P}$ differs from the original transition structure $P$, we calculate the Kullback-Leiber (KL) divergence between the two transition structures. The Kullback-Leiber divergence is a measure of how different a probability distribution is from a reference distribution, and is given by:
\begin{equation}
    D_{KL}(P || \hat{P}) = - \sum_i \pi_i \sum_j P_{ij} \; \log \, \frac{\hat{P}_{ij}}{P_{ij}} ,
    \label{eq:KL_div}
\end{equation}%
where $\vec{\pi}$ is the stationary probability distribution of the transition matrix $P$, obtained by solving $P \pi = \pi$. The KL-divergence between two quantities is always non-negative and attains the value zero if and only if $P = \hat{P}$. The larger the KL-divergence, the more the inferred network $\hat{P}$ differs from the original network. Hence, this quantity acts as a measure of the extent to which a network gets scrambled by the inaccuracies of human of learning---or in other words, how accurately the network structure is inferred.

\subsection{Null Models} \label{sec_null_models}
We aim to identify distinct features in the music networks that enable them to convey information effectively. To assess whether our observations are merely due to random chance or are instead a unique feature of our dataset, we compare our measurements on the real music networks with the following null network models \cite{newman2004finding, vavsa2022null}.
\begin{enumerate}
    \item \emph{Null networks with the same number of nodes and edges.} These are obtained by generating random networks with the same number of nodes and edges, and enable us to assess whether the quantity we have measured is to be expected merely based on network size.
    \item \emph{Degree-preserving null networks.} These are randomized networks of the same size, with the additional constraint that the in- and out-degrees of \emph{each} node in the network are preserved. Such networks are constructed by swapping edges between pairs of nodes in the network iteratively, such that the in- and out-degrees of each node are preserved but the connectivity (or topology) of the network is randomized. This class of null models enable us to evaluate the role that connectivity or topology plays in the quantity we are measuring. 
\end{enumerate}
We can generalize the degree-preserving null networks to weighted networks. We are interested in degree-preserving randomized networks since these keep the node-level entropies fixed and allow us to study the impact of topology on the quantities we are measuring. In the case of weighted networks, the node-level entropies are determined by the out-weights and out-degrees of the nodes. Hence, our procedure of swapping edges between pairs of nodes in the network still works since it preserved the out-weights of each node in addition to the in- and out-degrees. With these null models, we can benchmark the presence of the quantities we are interested in, and identify the role that the connectivity pattern or size plays. The code used to generate the null networks is available at \cite{SK_code}.

\subsection{Transitive Clustering Coefficient}
Along the lines of the clustering coefficient of a node \cite{smallworld, Szab2004Clustering}, we define the transitive clustering coefficient as a measure of the degree to which nodes in a directed network tend to form transitive relationships. The transitive clustering coefficient of a node $i$ (for an unweighted graph with no self loops) is given by:
\begin{equation}
    C_i^{T} =  \frac{\Delta^{T}_i}{ \, k^{\text{tot}}_i(k^{\text{tot}}_i-1)} ,
\end{equation}
where $\Delta^{T}_{i}$ denotes the number of transitive triangles that node $i$ is a part of and $k^{\text{tot}}_i$ is the total degree (in + out) of the node. The denominator simply counts the number of triangles that could exist within the neighborhood of node $i$.

\begin{figure}[h!]
    \centering
    \includegraphics[scale=0.5]{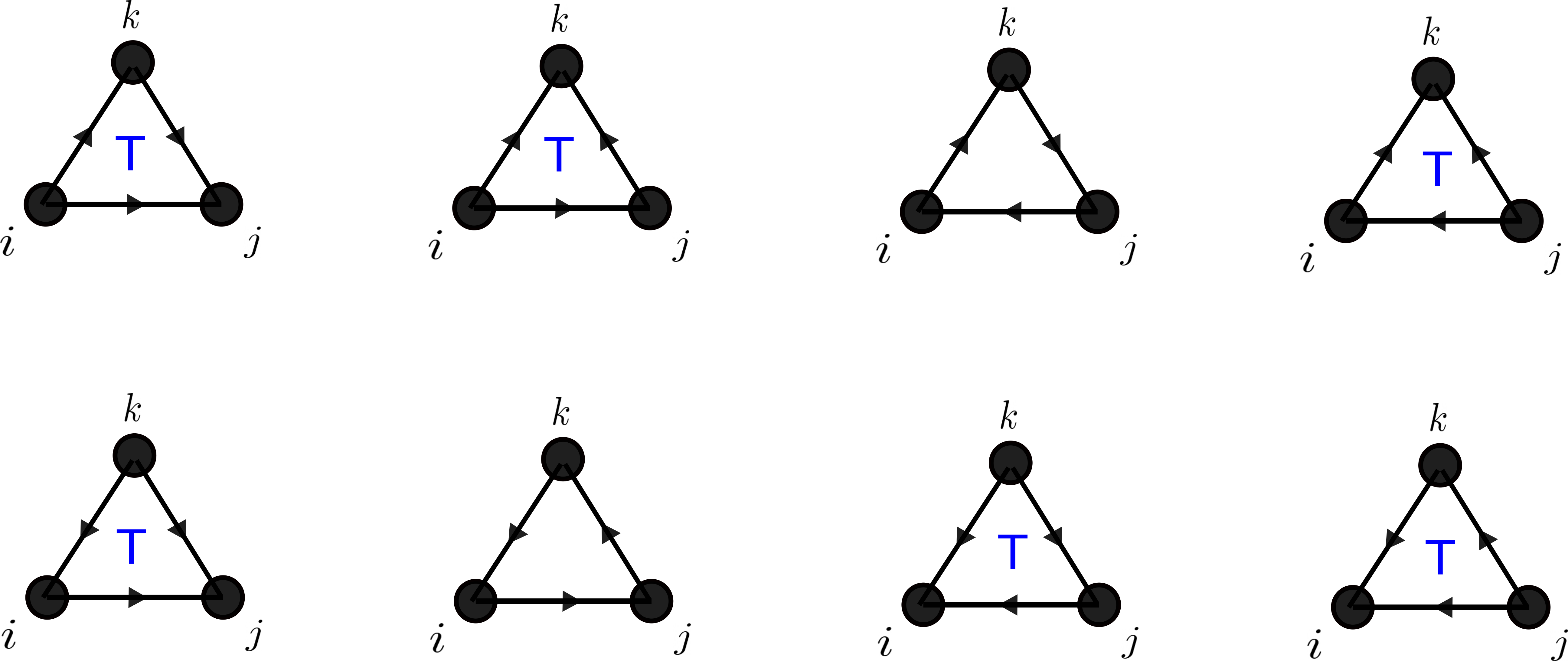}
    \caption{The 8 different possible triangles with node $i$ as a vertex in a directed graph. The triangles which represent transitive relationships are marked using the letter 'T'.}
    \label{fig:triangle_directed}
\end{figure}

The possible directed triangles involving node $i$ can be divided into two categories---those representing cyclic relationships and those representing transitive relationships (Fig. \ref{fig:triangle_directed}). The number of transitive triangles involving node $i$ that actually exist can be expressed in terms of the adjacency matrix of the graph $A$,  
\begin{equation}
    C_i^{T} = \frac{(A + A^{T})^3_{ii} - A^3_{ii} - (A^{T})_{ii}^{3}}{2 \, k^{\text{tot}}_i(k^{\text{tot}}_i-1)} .
\end{equation}
This expression counts a subset of the total number of triangles, and is a special case of the expression derived in Ref. \cite{fagiolo2007}. We will use this expression to measure the transitive clustering coefficient of each music networks (code available at \cite{SK_code}).

\section{Citation Diversity Statement}    
Recent work in several fields of science has identified a bias in citation practices such that papers from women and other minority scholars are under-cited relative to the number of such papers in the field \cite{mitchell2013gendered,dion2018gendered,caplar2017quantitative, maliniak2013gender, Dworkin2020.01.03.894378, bertolero2021racial, wang2021gendered, chatterjee2021gender, fulvio2021imbalance}. Here we sought to proactively consider choosing references that reflect the diversity of the field in thought, form of contribution, gender, race, ethnicity, and other factors. First, we obtained the predicted gender of the first and last author of each reference by using databases that store the probability of a first name being carried by a woman \cite{Dworkin2020.01.03.894378,zhou_dale_2020_3672110}. By this measure (and excluding self-citations to the first and last authors of our current paper), our references contain 9.37\% woman (first)/woman (last), 18.67\% man/woman, 19.29\% woman/man, and 52.67\% man/man. This method is limited in that a) names, pronouns, and social media profiles used to construct the databases may not, in every case, be indicative of gender identity and b) it cannot account for intersex, non-binary, or transgender people. Second, we obtained predicted racial/ethnic category of the first and last author of each reference by databases that store the probability of a first and last name being carried by an author of color \cite{ambekar2009name, sood2018predicting}. By this measure (and excluding self-citations), our references contain 11.79\% author of color (first)/author of color (last), 11.60\% white author/author of color, 16.05\% author of color/white author, and 60.56\% white author/white author. This method is limited in that a) names and Florida Voter Data to make the predictions may not be indicative of racial/ethnic identity, and b) it cannot account for Indigenous and mixed-race authors, or those who may face differential biases due to the ambiguous racialization or ethnicization of their names. We look forward to future work that could help us to better understand how to support equitable practices in science.

\section{Data and code availability}
The data and code used to perform the analyses in this paper are openly available at \url{https://github.com/SumanSKulkarni/Music_Networks}.

\section{Supplementary Information}

\subsection{Introduction}
In this Supplementary Information, we provide extended analysis and discussion to support the results presented in the main text. In Sec. \ref{S1}, we expand upon our analysis of the information content of Bach's music networks and how it relates to network structure. In Sec. \ref{S2}, we examine the transitive clustering coefficient more closely and study meso-scale features that might explain the differences observed across compositional forms.

\subsection{Information content} \label{S1}

\begin{figure*}[t!]
\centering
\includegraphics[width =\linewidth]{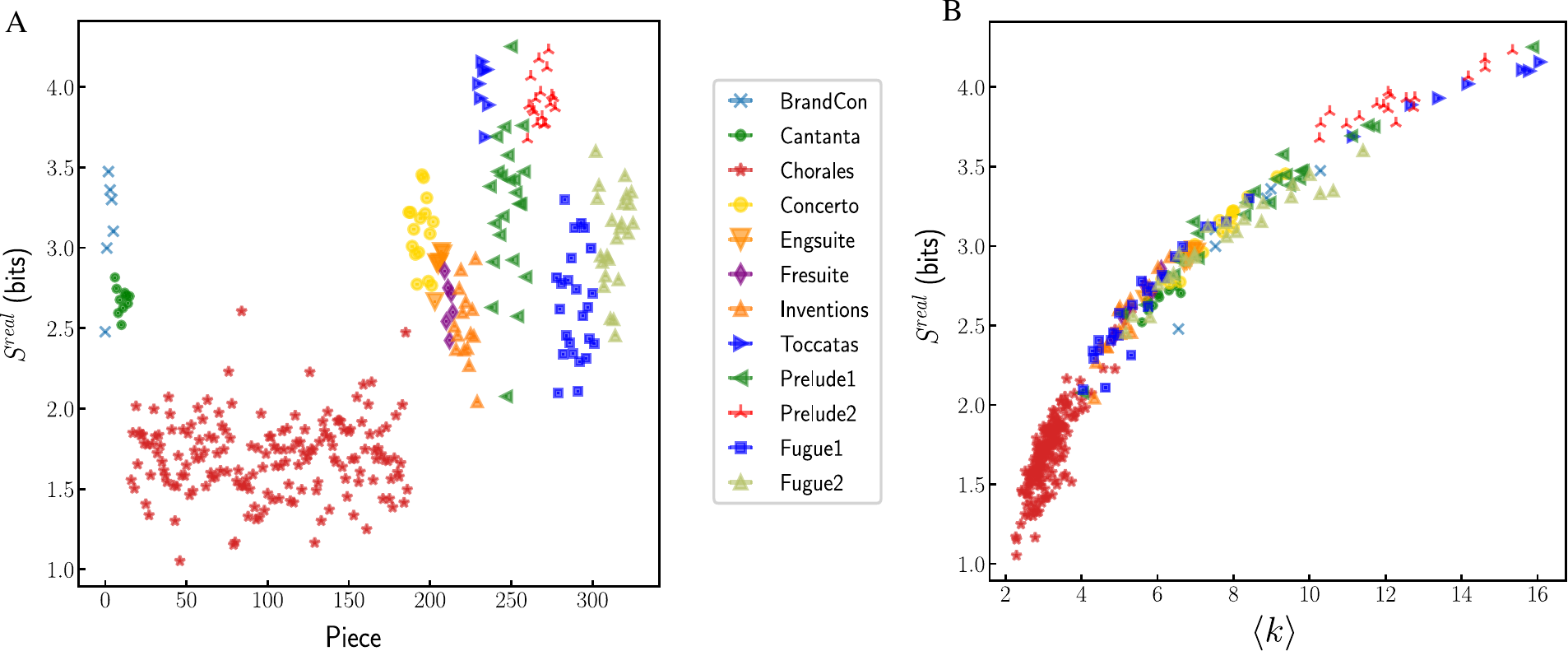}
\caption{\textbf{The entropy of Bach's music networks and its relation to the average degree of the network.} (A) The entropy of Bach's music networks ($S^{\text{real}}$) indexed by the pieces. (B) The entropy of Bach's music networks ($S^{\text{real}}$) as a function of the average degree of the network $\langle k \rangle$. Each data point in panels (A) and (B) represents a single piece. Colors and markers indicate the type of pieces, as shown in the legend.}
\label{fig:s_vs_k_1}
\end{figure*}

To better visualize the variation in information content among the musical compositions, we assign each piece an index number and plot the information entropy for each piece as a function of its index number (Fig. \ref{fig:s_vs_k_1}A). We observe here more clearly how different compositional forms tend to have pieces clustered together in their entropies. As reported in the main text, we find that the chorales have a markedly lower entropy than the rest of the compositions studied. In contrast, the toccatas and the second set of preludes have a much higher entropy. To relate the information entropy of the music networks to their structure, we compare their entropy to corresponding null networks (Fig. \ref{fig:unweighted_entropy}A and B in the main text), where we conclude that the information entropy is primarily determined by the degree distributions. In the case of undirected and unweighted networks, the network entropy depends upon the logarithm of the average degree of the network and the heterogeneity in the degree distribution (Eq. \ref{eq:entropy_taylor2}) to first and second order, respectively \cite{gomez2008,clynn:nat2020}. We now provide supplementary results that relate the information entropy of the music networks to their structure.

\subsection{Understanding the information entropy to first order: average degree} \label{S1_A}
\begin{figure*}[t!]
\centering
\includegraphics[width = 0.85\linewidth]{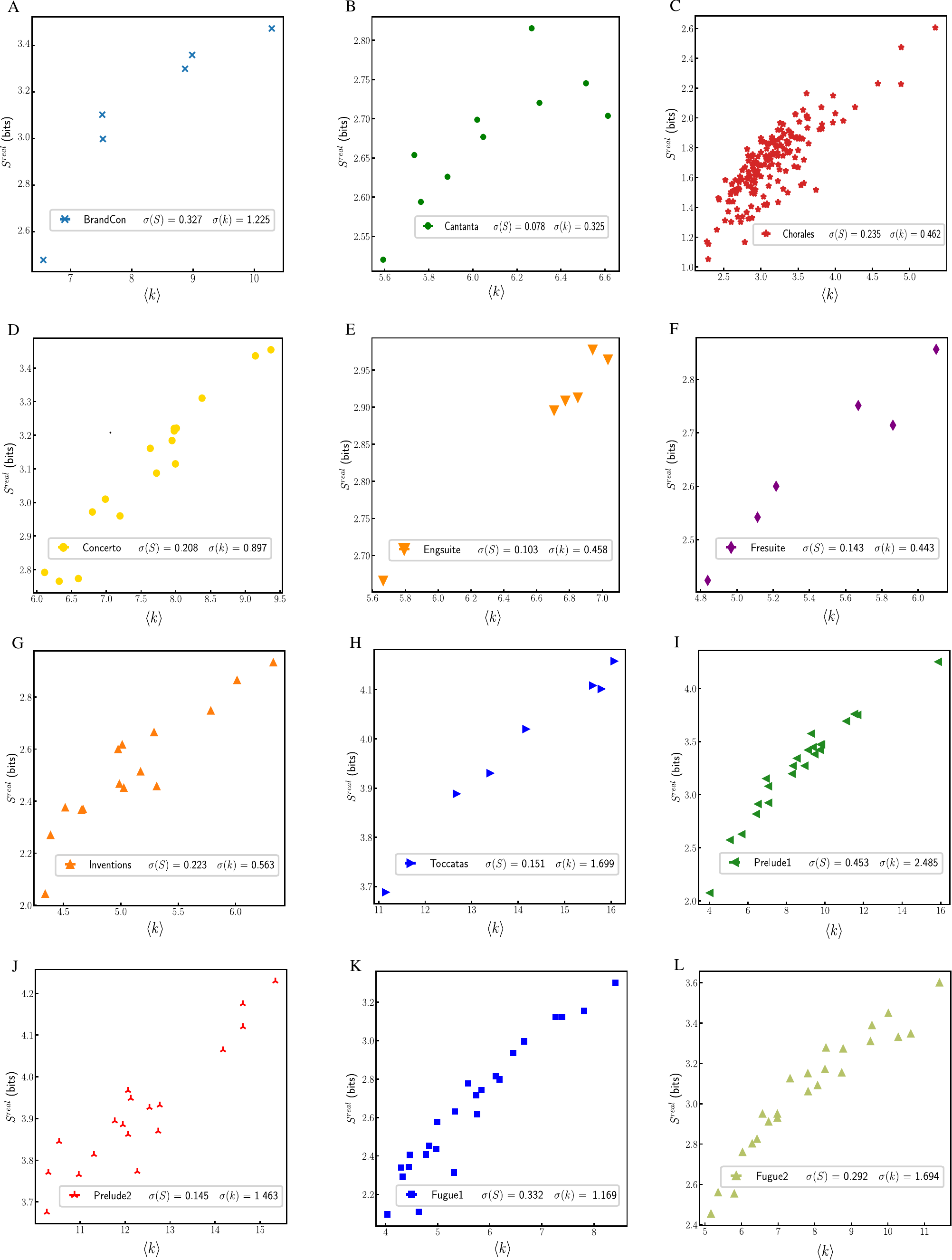}
\caption{\textbf{The relation between the information entropy and the average degree of the music networks plotted separately for each compositional form.} The entropy of Bach's music networks ($S^{\text{real}}$) plotted against the average degree of the network $\langle k \rangle$. Each data point represents a single piece. Colors and markers indicate the type of pieces, as shown in the legend.}
\label{fig:s_vs_k_2}
\end{figure*}

On plotting the information entropy of the music networks as a function of their average degree (Fig. \ref{fig:s_vs_k_1}B), we see that the differences in the information entropy of the compositional forms to first order arise due to differences in their average degrees. Although we observed in Fig. \ref{fig:s_vs_k_1}A that the compositional forms are clustered together in their entropy, it is clear that some pieces---such as the chorales, French suites, English suites, and cantatas---are more tightly clustered than the fugues and first set of preludes. These differences can be explained by the how much the average degrees vary across pieces. In Fig. \ref{fig:s_vs_k_2}, we plot the entropy of the music networks as a function of the average network degree, separately for each composition type. Additionally, we also report the standard deviation in the average degree of the pieces for each composition type.  Studying these plots, we observe that the English suites, French suites, and chorales (which clustered more tightly in their entropies) have tighter degree distributions, while the fugues (which are more spread out in their entropy) display more diverse average degrees.

\subsection{Understanding the information entropy to second order: degree heterogeneity} \label{S1_B}

\begin{figure*}[t!]
\centering
\includegraphics[width = 0.9\linewidth]{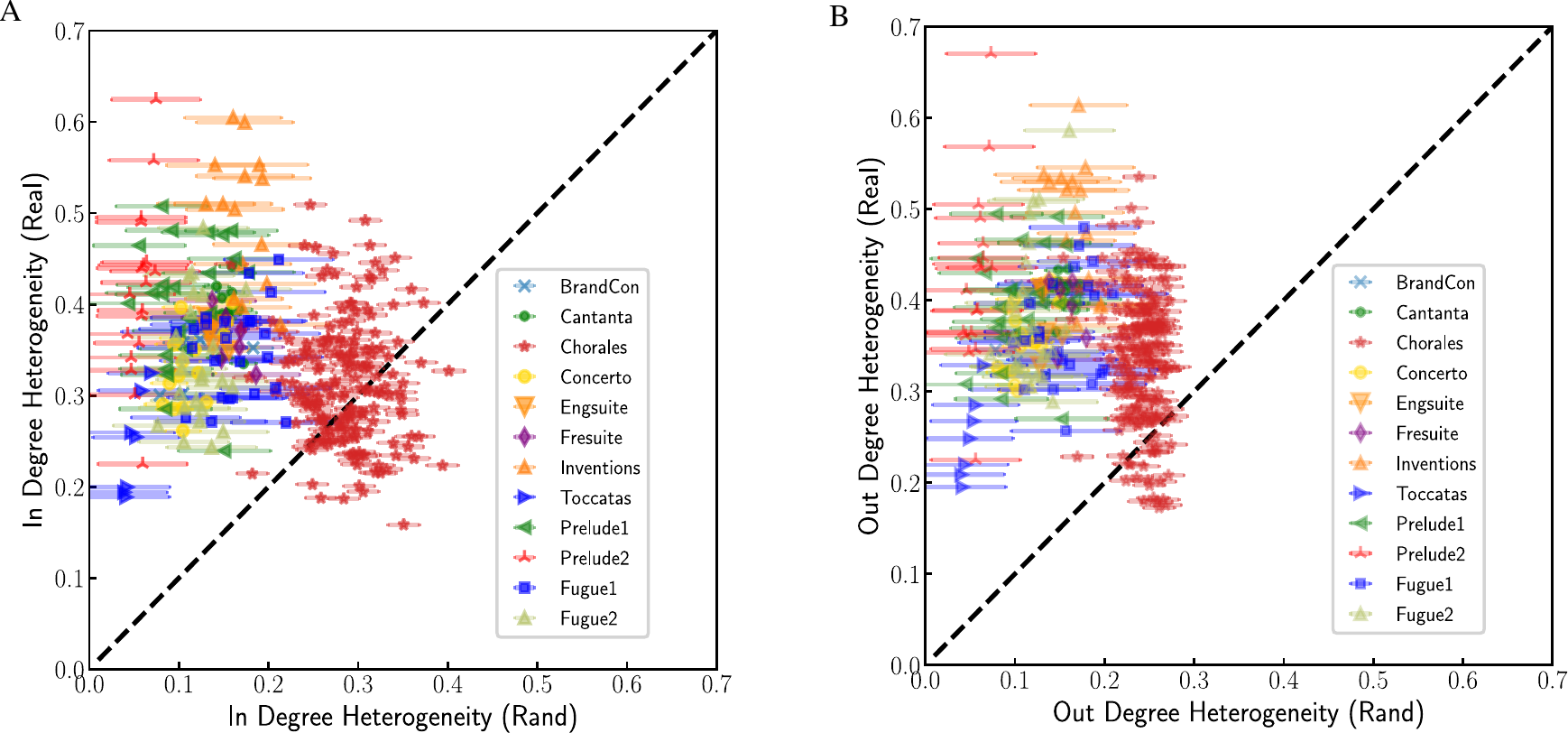}
\caption{\textbf{Comparing the heterogeneity of Bach's music networks to randomized null networks of the same size.} (A) The in-degree heterogeneity of the music networks compared with random networks of the same size. (B) The out-degree heterogeneity of the music networks compared with random networks of the same size. Each data point in panels (A) and (B) represents a single piece. Colors and markers indicate the type of pieces, as shown in the legend. For each random network, we report the in- and out- degree heterogeneity after averaging over 100 independent realizations. Error bars on the x-axis represent the standard error of the sample.}
\label{fig:s_vs_k_3}
\end{figure*}

In Fig. \ref{fig:unweighted_entropy}A of the main text, we observed that the entropy of the real music networks is larger than corresponding randomized null networks with the same number of nodes and edges. Since the average degree is the same for the two networks, we hypothesize that the differences arise due to higher in- and out-degree heterogeneity as per Eq. \ref{eq:entropy_taylor2}. To test our hypothesis, we compare the in- and out-degree heterogeneity of the music networks (calculated using Eq. \ref{eq:deg_het}) with their corresponding null networks in Fig. \ref{fig:s_vs_k_3}. In general, we observe that Bach's music networks are indeed more heterogeneous than expected from the random networks of the same size. This organization allows them to pack more information into their structure.

\begin{figure*}[t!]
\centering
\includegraphics[width = \linewidth]{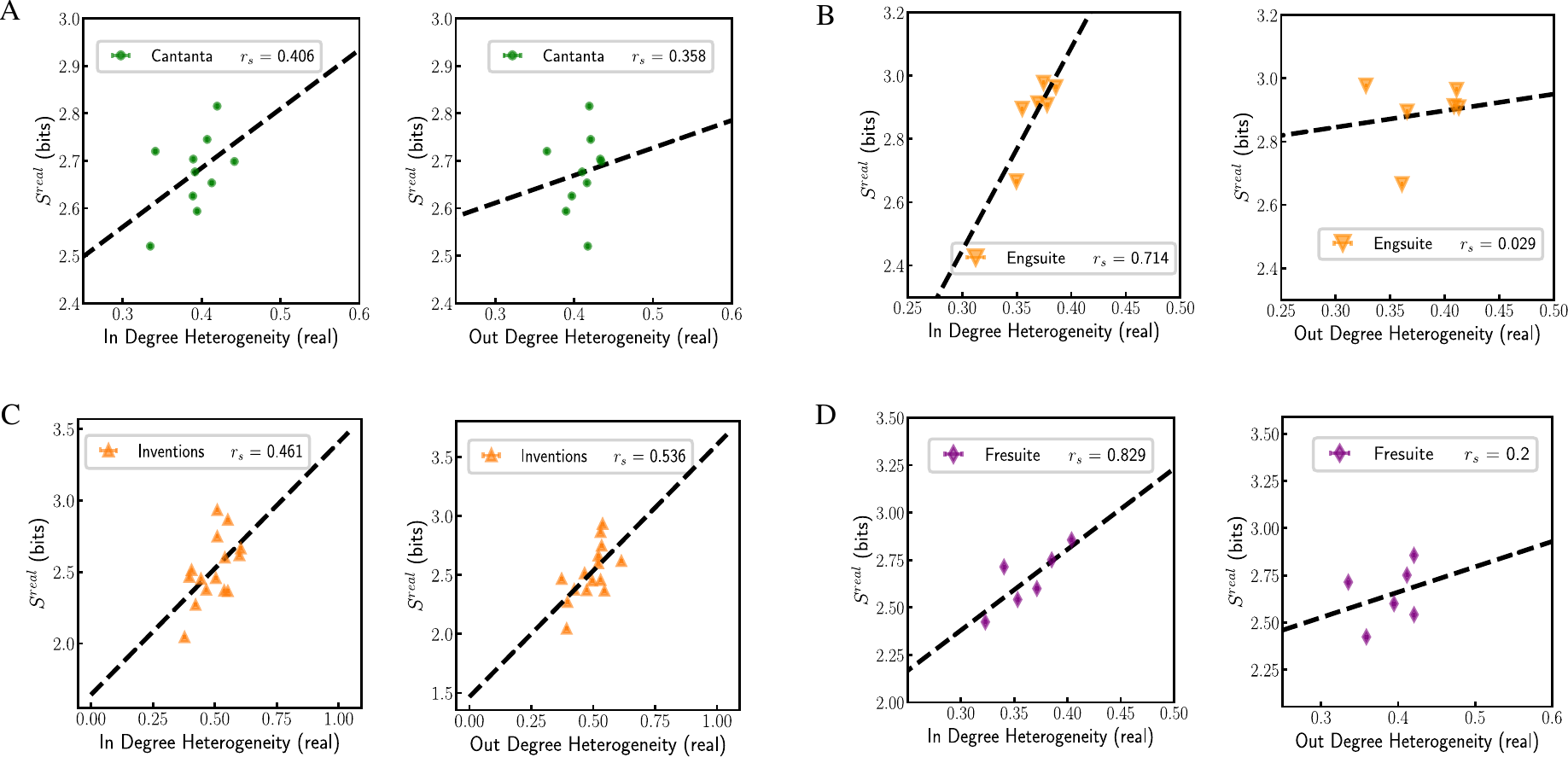}
\caption{\textbf{The relation between the information entropy of Bach's music networks and its degree heterogeneity.} The entropy of Bach's music networks ($S^{\text{real}}$) plotted against the network in- and out-degree heterogeneity. Each data point represents a single piece. Colors and markers indicate the type of pieces, as shown in the legend. The dotted line in each panel indicates the best linear fit, and the reported $r_{s}$ value is the Spearman correlation coefficient between the x- and y-axis variables.}
\label{fig:s_vs_k_4}
\end{figure*}

The heterogeneity in degrees can also explain the differences in entropies observed between pieces that are tightly clustered together in their entropy. As observed earlier, compositions such as the chorales, French suites, English suites, and cantatas have pieces that are clustered together in their average degree and consequentially, in their entropy. We expect that the differences observed among the pieces in each group can be explained by differences in their degree heterogeneity. In Fig. \ref{fig:s_vs_k_4} and Fig. \ref{fig:unweighted_entropy}C, we plot the entropies of the pieces that clustered together as a function of their in- and out-degree heterogeneity, and in general observe that the pieces with higher heterogeneity have a higher information entropy. However, we note that our sample size for most compositional forms is small and hence, we only report the chorales in the main text.

\subsection{Further analysis of the transitive clustering coefficient} \label{S2}
\begin{figure*}[t!]
\centering
\includegraphics[width = 0.8\linewidth]{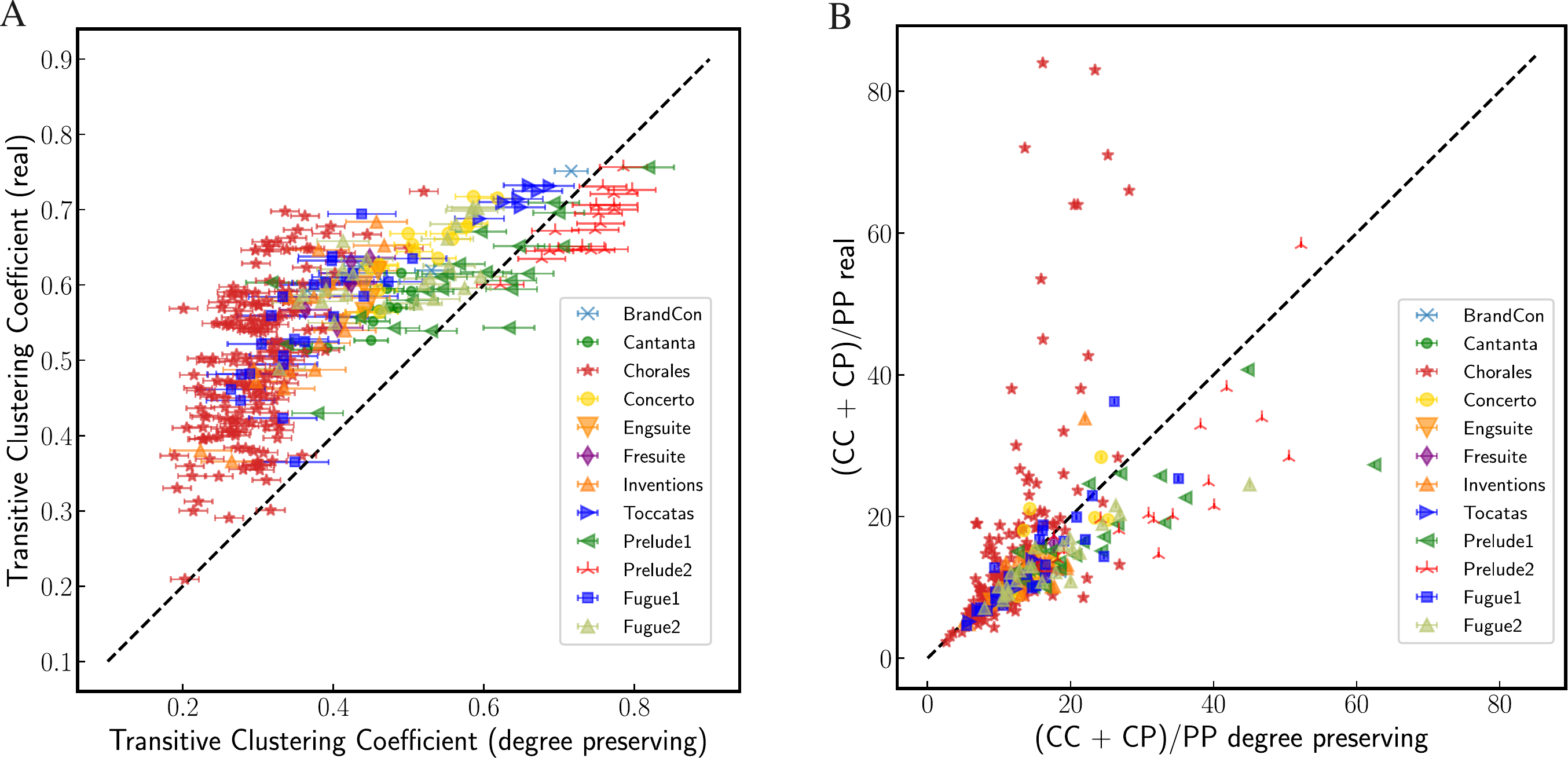}
\caption{\textbf{Core-periphery analysis of the music networks.} (A) The transitive clustering coefficient of the real music networks compared to null networks that preserve the in- and out-degree of each node. For the degree-preserving null networks, we report the average over 100 independent realizations, with error bars denoting the standard error of the sample. (B) The ratio of the number of core-core (CC) edges and core-periphery (CP) edges to the number of periphery-periphery (PP) edges in the real music networks compared to degree-preserving null networks. For the degree-preserving null networks, we report the average value computed over 100 independent random graphs. In both panels, the dotted line indicates the line $y=x$. Colors and markers indicate the type of piece, as shown in the legend.}
\label{fig:coreness}
\end{figure*}

In our analysis of the discrepancies between the actual and perceived information content of note transitions in Bach's musical compositions, we found that these discrepancies were primarily driven by the presence of transitive triangular clusters. These transitive triangular clusters tend to bring the inferred network closer to the actual network, making the network more learnable. As shown in Fig. \ref{fig:coreness}A, the real (unweighted) music networks tend to have a higher transitive clustering coefficient than random networks that preserve the degree of each node, indicating that this is a distinct feature of the music networks that is not merely due to coincidence. The data in Fig. \ref{fig:coreness}A has a striking shape, which we elaborate on and analyze in this section. First we observe that the chorale pieces tend to have a higher transitive clustering coefficient than expected from networks of their same size and degree distribution. Second, although the preludes have a higher transitive clustering coefficient than other compositional forms, the value was still lower than expected from networks of their same size and degree distribution. Indeed, by examining only the x-axis, we notice that the null networks corresponding to the preludes have a higher transitive clustering coefficient than the null networks corresponding to chorales. However, by examining the y-axis, we see that the deviation between the real chorales and the prelude networks are not that pronounced. We hypothesize that these differences might be due to the presence of mesoscale features in the networks, such as core-periphery structure.

\subsubsection{Core-periphery structure}
Core-periphery structure in a network refers to the presence of two components: a tightly connected ``core'' and a sparsely connected ``periphery. The core consists of nodes which are well-connected to each other and to the periphery, while the nodes in the periphery are sparsely connected to one another and to the nodes in the core \cite{rombach2014core, borgatti2000}. We hypothesize that the presence of a relatively larger core might explain why the chorales have a higher clustering coefficient than expected given their size and degree. Similarly, a smaller than expected core for the preludes might explain why their clustering coefficient was lower than expected from networks of the same size and degree distribution. Since the core consists of nodes that are well-connected to themselves and the periphery, if there are a larger number of edges occurring within the core and between the core and periphery than between the periphery nodes, it is likely that these edges will form the clusters that we are interested in. We denote the edges between two nodes that belong to the core by core-core (CC), those between nodes that belong to the periphery by periphery-periphery (PP), and those between the nodes in the core and the nodes in the periphery by core-periphery (CP). 

To test our hypothesis, we compute the core-periphery structure for each music network using the method described by Borgatti and Everett \cite{borgatti2000}. We then compute the ratio of the sum of the number of core-core (CC) edges and core-periphery (CP) edges to the number of periphery-periphery (PP) edges for each network. To understand this ratio, we compare it to corresponding degree-preserving null networks (Fig. \ref{fig:coreness}B). Strikingly, we observe that the chorales have a higher fraction of edges that are within or emanating from the core than expected from their corresponding null networks. The preludes are at the other end, and have a lower fraction of edges that are within or emanating from the core than expected from their corresponding null networks. This pattern of findings suggests that the chorales have a more pronounced core-periphery structure than expected by chance, while the preludes have a less pronounced core-periphery structure than expected. Although the preludes still have a slightly higher transitive clustering coefficient than the other pieces, the differences are not as pronounced as one would expect because of these differences in their core-periphery structure.

By performing this additional analysis, we provide an example of how the music networks display interesting meso-scale structures that differ from one compositional form to another, resulting in differences in how their network structure is perceived.

\newpage
\clearpage

\bibliography{bach_refs_arxiv}

\begin{thebibliography}{94}%
\makeatletter
\providecommand \@ifxundefined [1]{%
 \@ifx{#1\undefined}
}%
\providecommand \@ifnum [1]{%
 \ifnum #1\expandafter \@firstoftwo
 \else \expandafter \@secondoftwo
 \fi
}%
\providecommand \@ifx [1]{%
 \ifx #1\expandafter \@firstoftwo
 \else \expandafter \@secondoftwo
 \fi
}%
\providecommand \natexlab [1]{#1}%
\providecommand \enquote  [1]{``#1''}%
\providecommand \bibnamefont  [1]{#1}%
\providecommand \bibfnamefont [1]{#1}%
\providecommand \citenamefont [1]{#1}%
\providecommand \href@noop [0]{\@secondoftwo}%
\providecommand \href [0]{\begingroup \@sanitize@url \@href}%
\providecommand \@href[1]{\@@startlink{#1}\@@href}%
\providecommand \@@href[1]{\endgroup#1\@@endlink}%
\providecommand \@sanitize@url [0]{\catcode `\\12\catcode `\$12\catcode
  `\&12\catcode `\#12\catcode `\^12\catcode `\_12\catcode `\%12\relax}%
\providecommand \@@startlink[1]{}%
\providecommand \@@endlink[0]{}%
\providecommand \url  [0]{\begingroup\@sanitize@url \@url }%
\providecommand \@url [1]{\endgroup\@href {#1}{\urlprefix }}%
\providecommand \urlprefix  [0]{URL }%
\providecommand \Eprint [0]{\href }%
\providecommand \doibase [0]{http://dx.doi.org/}%
\providecommand \selectlanguage [0]{\@gobble}%
\providecommand \bibinfo  [0]{\@secondoftwo}%
\providecommand \bibfield  [0]{\@secondoftwo}%
\providecommand \translation [1]{[#1]}%
\providecommand \BibitemOpen [0]{}%
\providecommand \bibitemStop [0]{}%
\providecommand \bibitemNoStop [0]{.\EOS\space}%
\providecommand \EOS [0]{\spacefactor3000\relax}%
\providecommand \BibitemShut  [1]{\csname bibitem#1\endcsname}%
\let\auto@bib@innerbib\@empty
%</preamble>
\bibitem [{\citenamefont {Mithen}(2005)}]{mithen2005the}%
  \BibitemOpen
  \bibfield  {author} {\bibinfo {author} {\bibfnamefont {S.~J.}\ \bibnamefont
  {Mithen}},\ }\href@noop {} {\emph {\bibinfo {title} {The Singing
  Neanderthals: the origins of music, language, mind and body}}}\ (\bibinfo
  {publisher} {Harvard University Press},\ \bibinfo {year} {2005})\BibitemShut
  {NoStop}%
\bibitem [{\citenamefont {Welch}\ \emph {et~al.}(2020)\citenamefont {Welch},
  \citenamefont {Biasutti}, \citenamefont {MacRitchie}, \citenamefont
  {McPherson},\ and\ \citenamefont {Himonides}}]{welch2020impact}%
  \BibitemOpen
  \bibfield  {author} {\bibinfo {author} {\bibfnamefont {G.~F.}\ \bibnamefont
  {Welch}}, \bibinfo {author} {\bibfnamefont {M.}~\bibnamefont {Biasutti}},
  \bibinfo {author} {\bibfnamefont {J.}~\bibnamefont {MacRitchie}}, \bibinfo
  {author} {\bibfnamefont {G.~E.}\ \bibnamefont {McPherson}}, \ and\ \bibinfo
  {author} {\bibfnamefont {E.}~\bibnamefont {Himonides}},\ }\href@noop {}
  {\enquote {\bibinfo {title} {The impact of music on human development and
  well-being},}\ } (\bibinfo {year} {2020})\BibitemShut {NoStop}%
\bibitem [{\citenamefont {Cross}(2001)}]{cross2001music}%
  \BibitemOpen
  \bibfield  {author} {\bibinfo {author} {\bibfnamefont {I.}~\bibnamefont
  {Cross}},\ }\bibfield  {title} {\enquote {\bibinfo {title} {Music, cognition,
  culture, and evolution},}\ }\href@noop {} {\bibfield  {journal} {\bibinfo
  {journal} {Annals of the New York Academy of sciences}\ }\textbf {\bibinfo
  {volume} {930}},\ \bibinfo {pages} {28--42} (\bibinfo {year}
  {2001})}\BibitemShut {NoStop}%
\bibitem [{\citenamefont {McClary}(1997)}]{McClary1997the}%
  \BibitemOpen
  \bibfield  {author} {\bibinfo {author} {\bibfnamefont {S.}~\bibnamefont
  {McClary}},\ }\bibfield  {title} {\enquote {\bibinfo {title} {The impromptu
  that trod on a loaf: Or how music tells stories},}\ }\href@noop {} {\bibfield
   {journal} {\bibinfo  {journal} {Narrative}\ }\textbf {\bibinfo {volume}
  {5}},\ \bibinfo {pages} {20–35} (\bibinfo {year} {1997})}\BibitemShut
  {NoStop}%
\bibitem [{\citenamefont {Glennie}\ \emph {et~al.}(2005)\citenamefont
  {Glennie}, \citenamefont {Donald} \emph {et~al.}}]{glennie2005musical}%
  \BibitemOpen
  \bibfield  {author} {\bibinfo {author} {\bibfnamefont {E.}~\bibnamefont
  {Glennie}}, \bibinfo {author} {\bibfnamefont {E.~Mac}\ \bibnamefont
  {Donald}},  \emph {et~al.},\ }\href@noop {} {\emph {\bibinfo {title} {Musical
  communication}}}\ (\bibinfo  {publisher} {Oxford University Press on
  Demand},\ \bibinfo {year} {2005})\BibitemShut {NoStop}%
\bibitem [{\citenamefont {Scherer}\ and\ \citenamefont
  {Coutinho}(2013)}]{scherer2013music}%
  \BibitemOpen
  \bibfield  {author} {\bibinfo {author} {\bibfnamefont {K.~R}\ \bibnamefont
  {Scherer}}\ and\ \bibinfo {author} {\bibfnamefont {E.}~\bibnamefont
  {Coutinho}},\ }\bibfield  {title} {\enquote {\bibinfo {title} {How music
  creates emotion: a multifactorial process approach},}\ }\href@noop {}
  {\bibfield  {journal} {\bibinfo  {journal} {The emotional power of music:
  Multidisciplinary perspectives on musical arousal, expression, and social
  control}\ ,\ \bibinfo {pages} {121--145}} (\bibinfo {year}
  {2013})}\BibitemShut {NoStop}%
\bibitem [{\citenamefont {Koelsch}(2014)}]{koelsch2014}%
  \BibitemOpen
  \bibfield  {author} {\bibinfo {author} {\bibfnamefont {S.}~\bibnamefont
  {Koelsch}},\ }\bibfield  {title} {\enquote {\bibinfo {title} {Brain
  correlates of music-evoked emotions},}\ }\href@noop {} {\bibfield  {journal}
  {\bibinfo  {journal} {Nature Reviews Neuroscience}\ }\textbf {\bibinfo
  {volume} {15}},\ \bibinfo {pages} {170--180} (\bibinfo {year}
  {2014})}\BibitemShut {NoStop}%
\bibitem [{\citenamefont {Blood}\ and\ \citenamefont
  {Zatorre}(2001)}]{blood2001}%
  \BibitemOpen
  \bibfield  {author} {\bibinfo {author} {\bibfnamefont {A.~J.}\ \bibnamefont
  {Blood}}\ and\ \bibinfo {author} {\bibfnamefont {R.~J.}\ \bibnamefont
  {Zatorre}},\ }\bibfield  {title} {\enquote {\bibinfo {title} {Intensely
  pleasurable responses to music correlate with activity in brain regions
  implicated in reward and emotion},}\ }\href@noop {} {\bibfield  {journal}
  {\bibinfo  {journal} {Proceedings of the National Academy of Sciences}\
  }\textbf {\bibinfo {volume} {98}},\ \bibinfo {pages} {11818--11823} (\bibinfo
  {year} {2001})}\BibitemShut {NoStop}%
\bibitem [{\citenamefont {Huron}(2006)}]{huron2006sweet}%
  \BibitemOpen
  \bibfield  {author} {\bibinfo {author} {\bibfnamefont {D.}~\bibnamefont
  {Huron}},\ }\href@noop {} {\emph {\bibinfo {title} {Sweet Anticipation: Music
  and the Psychology of Expectation}}}\ (\bibinfo  {publisher} {The MIT
  Press},\ \bibinfo {year} {2006})\BibitemShut {NoStop}%
\bibitem [{\citenamefont {Tillmann}\ \emph {et~al.}(2014)\citenamefont
  {Tillmann}, \citenamefont {Poulin-Charronnat},\ and\ \citenamefont
  {Bigand}}]{tillmann2014}%
  \BibitemOpen
  \bibfield  {author} {\bibinfo {author} {\bibfnamefont {B.}~\bibnamefont
  {Tillmann}}, \bibinfo {author} {\bibfnamefont {B.}~\bibnamefont
  {Poulin-Charronnat}}, \ and\ \bibinfo {author} {\bibfnamefont
  {E.}~\bibnamefont {Bigand}},\ }\bibfield  {title} {\enquote {\bibinfo {title}
  {The role of expectation in music: from the score to emotions and the
  brain},}\ }\href@noop {} {\bibfield  {journal} {\bibinfo  {journal} {WIREs
  Cognitive Science}\ }\textbf {\bibinfo {volume} {5}},\ \bibinfo {pages}
  {105--113} (\bibinfo {year} {2014})}\BibitemShut {NoStop}%
\bibitem [{\citenamefont {Pearce}\ and\ \citenamefont
  {Wiggins}(2012)}]{pearce2012auditory}%
  \BibitemOpen
  \bibfield  {author} {\bibinfo {author} {\bibfnamefont {M.~T.}\ \bibnamefont
  {Pearce}}\ and\ \bibinfo {author} {\bibfnamefont {G.~A}\ \bibnamefont
  {Wiggins}},\ }\bibfield  {title} {\enquote {\bibinfo {title} {Auditory
  expectation: the information dynamics of music perception and cognition},}\
  }\href@noop {} {\bibfield  {journal} {\bibinfo  {journal} {Topics in
  cognitive science}\ }\textbf {\bibinfo {volume} {4}},\ \bibinfo {pages}
  {625--652} (\bibinfo {year} {2012})}\BibitemShut {NoStop}%
\bibitem [{\citenamefont {Meyer}(1956)}]{meyer1956emotion}%
  \BibitemOpen
  \bibfield  {author} {\bibinfo {author} {\bibfnamefont {L.~B.}\ \bibnamefont
  {Meyer}},\ }\href@noop {} {\emph {\bibinfo {title} {Emotion and Meaning in
  Music}}},\ ACLS Humanities E-Book\ (\bibinfo  {publisher} {University of
  Chicago Press},\ \bibinfo {year} {1956})\BibitemShut {NoStop}%
\bibitem [{\citenamefont {Ferretti}(2017)}]{ferretti2017modeling}%
  \BibitemOpen
  \bibfield  {author} {\bibinfo {author} {\bibfnamefont {S.}~\bibnamefont
  {Ferretti}},\ }\bibfield  {title} {\enquote {\bibinfo {title} {On the
  modeling of musical solos as complex networks},}\ }\href@noop {} {\bibfield
  {journal} {\bibinfo  {journal} {Information Sciences}\ }\textbf {\bibinfo
  {volume} {375}} (\bibinfo {year} {2017})}\BibitemShut {NoStop}%
\bibitem [{\citenamefont {Ferretti}(2018)}]{ferretti2018complex}%
  \BibitemOpen
  \bibfield  {author} {\bibinfo {author} {\bibfnamefont {S.}~\bibnamefont
  {Ferretti}},\ }\bibfield  {title} {\enquote {\bibinfo {title} {On the complex
  network structure of musical pieces: analysis of some use cases from
  different music genres},}\ }\href@noop {} {\bibfield  {journal} {\bibinfo
  {journal} {Multimedia Tools and Applications}\ }\textbf {\bibinfo {volume}
  {77}},\ \bibinfo {pages} {16003--16029} (\bibinfo {year} {2018})}\BibitemShut
  {NoStop}%
\bibitem [{\citenamefont {Liu}\ \emph {et~al.}(2010)\citenamefont {Liu},
  \citenamefont {Chi},\ and\ \citenamefont {Small}}]{Liu2010}%
  \BibitemOpen
  \bibfield  {author} {\bibinfo {author} {\bibfnamefont {X.~F.}\ \bibnamefont
  {Liu}}, \bibinfo {author} {\bibfnamefont {K.~T.}\ \bibnamefont {Chi}}, \ and\
  \bibinfo {author} {\bibfnamefont {M.}~\bibnamefont {Small}},\ }\bibfield
  {title} {\enquote {\bibinfo {title} {Complex network structure of musical
  compositions: Algorithmic generation of appealing music},}\ }\href@noop {}
  {\bibfield  {journal} {\bibinfo  {journal} {Physica A: Statistical Mechanics
  and its Applications}\ }\textbf {\bibinfo {volume} {389}},\ \bibinfo {pages}
  {126--132} (\bibinfo {year} {2010})}\BibitemShut {NoStop}%
\bibitem [{\citenamefont {B.~Nardelli}(2021)}]{buongiorno2021musicntwrk}%
  \BibitemOpen
  \bibfield  {author} {\bibinfo {author} {\bibfnamefont {M.}~\bibnamefont
  {B.~Nardelli}},\ }\bibfield  {title} {\enquote {\bibinfo {title} {Musicntwrk:
  data tools for music theory, analysis and composition},}\ }in\ \href@noop {}
  {\emph {\bibinfo {booktitle} {Perception, Representations, Image, Sound,
  Music: 14th International Symposium, CMMR 2019, Marseille, France, October
  14--18, 2019, Revised Selected Papers 14}}}\ (\bibinfo {organization}
  {Springer},\ \bibinfo {year} {2021})\ pp.\ \bibinfo {pages}
  {190--215}\BibitemShut {NoStop}%
\bibitem [{\citenamefont {{C. W. Lynn, L. Papadopoulos, A. E. Kahn, and D. S.
  Bassett}}(2020)}]{clynn:nat2020}%
  \BibitemOpen
  \bibfield  {author} {\bibinfo {author} {\bibnamefont {{C. W. Lynn, L.
  Papadopoulos, A. E. Kahn, and D. S. Bassett}}},\ }\bibfield  {title}
  {\enquote {\bibinfo {title} {{Human information processing in complex
  networks}},}\ }\href@noop {} {\bibfield  {journal} {\bibinfo  {journal}
  {Nature Physics}\ }\textbf {\bibinfo {volume} {16}},\ \bibinfo {pages} {965}
  (\bibinfo {year} {2020})}\BibitemShut {NoStop}%
\bibitem [{\citenamefont {{C. E. Shannon}}(1948)}]{shannon}%
  \BibitemOpen
  \bibfield  {author} {\bibinfo {author} {\bibnamefont {{C. E. Shannon}}},\
  }\bibfield  {title} {\enquote {\bibinfo {title} {{A mathematical theory of
  communication}},}\ }\href@noop {} {\bibfield  {journal} {\bibinfo  {journal}
  {The Bell System Technical Journal}\ }\textbf {\bibinfo {volume} {27}},\
  \bibinfo {pages} {379--423} (\bibinfo {year} {1948})}\BibitemShut {NoStop}%
\bibitem [{\citenamefont {Piantadosi}\ \emph {et~al.}(2011)\citenamefont
  {Piantadosi}, \citenamefont {Tily},\ and\ \citenamefont
  {Gibson}}]{piantadosi2011}%
  \BibitemOpen
  \bibfield  {author} {\bibinfo {author} {\bibfnamefont {S.~T.}\ \bibnamefont
  {Piantadosi}}, \bibinfo {author} {\bibfnamefont {H.}~\bibnamefont {Tily}}, \
  and\ \bibinfo {author} {\bibfnamefont {E.}~\bibnamefont {Gibson}},\
  }\bibfield  {title} {\enquote {\bibinfo {title} {Word lengths are optimized
  for efficient communication},}\ }\href@noop {} {\bibfield  {journal}
  {\bibinfo  {journal} {Proceedings of the National Academy of Sciences}\
  }\textbf {\bibinfo {volume} {108}},\ \bibinfo {pages} {3526--3529} (\bibinfo
  {year} {2011})}\BibitemShut {NoStop}%
\bibitem [{\citenamefont {Plotkin}\ and\ \citenamefont
  {Nowak}(2000)}]{plotkin2000language}%
  \BibitemOpen
  \bibfield  {author} {\bibinfo {author} {\bibfnamefont {J.~B.}\ \bibnamefont
  {Plotkin}}\ and\ \bibinfo {author} {\bibfnamefont {M.~A.}\ \bibnamefont
  {Nowak}},\ }\bibfield  {title} {\enquote {\bibinfo {title} {Language
  evolution and information theory},}\ }\href@noop {} {\bibfield  {journal}
  {\bibinfo  {journal} {Journal of Theoretical Biology}\ }\textbf {\bibinfo
  {volume} {205}},\ \bibinfo {pages} {147--159} (\bibinfo {year}
  {2000})}\BibitemShut {NoStop}%
\bibitem [{\citenamefont {Eckmann}\ \emph {et~al.}(2004)\citenamefont
  {Eckmann}, \citenamefont {Moses},\ and\ \citenamefont {Sergi}}]{eckmann2004}%
  \BibitemOpen
  \bibfield  {author} {\bibinfo {author} {\bibfnamefont {J.-P.}\ \bibnamefont
  {Eckmann}}, \bibinfo {author} {\bibfnamefont {E.}~\bibnamefont {Moses}}, \
  and\ \bibinfo {author} {\bibfnamefont {D.}~\bibnamefont {Sergi}},\ }\bibfield
   {title} {\enquote {\bibinfo {title} {Entropy of dialogues creates coherent
  structures in e-mail traffic},}\ }\href@noop {} {\bibfield  {journal}
  {\bibinfo  {journal} {Proceedings of the National Academy of Sciences}\
  }\textbf {\bibinfo {volume} {101}},\ \bibinfo {pages} {14333--14337}
  (\bibinfo {year} {2004})}\BibitemShut {NoStop}%
\bibitem [{\citenamefont {Zhao}\ \emph {et~al.}(2011)\citenamefont {Zhao},
  \citenamefont {Karsai},\ and\ \citenamefont {Bianconi}}]{zhao2011}%
  \BibitemOpen
  \bibfield  {author} {\bibinfo {author} {\bibfnamefont {K.}~\bibnamefont
  {Zhao}}, \bibinfo {author} {\bibfnamefont {M.}~\bibnamefont {Karsai}}, \ and\
  \bibinfo {author} {\bibfnamefont {G.}~\bibnamefont {Bianconi}},\ }\bibfield
  {title} {\enquote {\bibinfo {title} {Entropy of dynamical social networks},}\
  }\href@noop {} {\bibfield  {journal} {\bibinfo  {journal} {PLOS ONE}\ ,\
  \bibinfo {pages} {1--7}} (\bibinfo {year} {2011})}\BibitemShut {NoStop}%
\bibitem [{\citenamefont {Rosvall}\ \emph {et~al.}(2005)\citenamefont
  {Rosvall}, \citenamefont {Trusina}, \citenamefont {Minnhagen},\ and\
  \citenamefont {Sneppen}}]{rosvall2005}%
  \BibitemOpen
  \bibfield  {author} {\bibinfo {author} {\bibfnamefont {M.}~\bibnamefont
  {Rosvall}}, \bibinfo {author} {\bibfnamefont {A.}~\bibnamefont {Trusina}},
  \bibinfo {author} {\bibfnamefont {P.}~\bibnamefont {Minnhagen}}, \ and\
  \bibinfo {author} {\bibfnamefont {K.}~\bibnamefont {Sneppen}},\ }\bibfield
  {title} {\enquote {\bibinfo {title} {Networks and cities: An information
  perspective},}\ }\href@noop {} {\bibfield  {journal} {\bibinfo  {journal}
  {Phys. Rev. Lett.}\ }\textbf {\bibinfo {volume} {94}},\ \bibinfo {pages}
  {028701} (\bibinfo {year} {2005})}\BibitemShut {NoStop}%
\bibitem [{\citenamefont {Cohen}(1962)}]{cohen1962information}%
  \BibitemOpen
  \bibfield  {author} {\bibinfo {author} {\bibfnamefont {J.~E.}\ \bibnamefont
  {Cohen}},\ }\bibfield  {title} {\enquote {\bibinfo {title} {Information
  theory and music},}\ }\href@noop {} {\bibfield  {journal} {\bibinfo
  {journal} {Behavioral Science}\ }\textbf {\bibinfo {volume} {7}},\ \bibinfo
  {pages} {137--163} (\bibinfo {year} {1962})}\BibitemShut {NoStop}%
\bibitem [{\citenamefont {Hiller}\ and\ \citenamefont
  {Bean}(1966)}]{hiller1966information}%
  \BibitemOpen
  \bibfield  {author} {\bibinfo {author} {\bibfnamefont {L.}~\bibnamefont
  {Hiller}}\ and\ \bibinfo {author} {\bibfnamefont {C.}~\bibnamefont {Bean}},\
  }\bibfield  {title} {\enquote {\bibinfo {title} {Information theory analyses
  of four sonata expositions},}\ }\href@noop {} {\bibfield  {journal} {\bibinfo
   {journal} {Journal of Music Theory}\ }\textbf {\bibinfo {volume} {10}},\
  \bibinfo {pages} {96--137} (\bibinfo {year} {1966})}\BibitemShut {NoStop}%
\bibitem [{\citenamefont {Gomez}\ \emph {et~al.}(2014)\citenamefont {Gomez},
  \citenamefont {Lorimer},\ and\ \citenamefont {Stoop}}]{gomez2014}%
  \BibitemOpen
  \bibfield  {author} {\bibinfo {author} {\bibfnamefont {F.}~\bibnamefont
  {Gomez}}, \bibinfo {author} {\bibfnamefont {T.}~\bibnamefont {Lorimer}}, \
  and\ \bibinfo {author} {\bibfnamefont {R.}~\bibnamefont {Stoop}},\ }\bibfield
   {title} {\enquote {\bibinfo {title} {Complex networks of harmonic structure
  in classical music},}\ }in\ \href@noop {} {\emph {\bibinfo {booktitle}
  {Nonlinear Dynamics of Electronic Systems}}}\ (\bibinfo  {publisher}
  {Springer International Publishing},\ \bibinfo {address} {Cham},\ \bibinfo
  {year} {2014})\ pp.\ \bibinfo {pages} {262--269}\BibitemShut {NoStop}%
\bibitem [{\citenamefont {Boon}\ \emph {et~al.}(1990)\citenamefont {Boon},
  \citenamefont {Noullez},\ and\ \citenamefont {Mommen}}]{Boon1990}%
  \BibitemOpen
  \bibfield  {author} {\bibinfo {author} {\bibfnamefont {J.‐P.}\ \bibnamefont
  {Boon}}, \bibinfo {author} {\bibfnamefont {A.}~\bibnamefont {Noullez}}, \
  and\ \bibinfo {author} {\bibfnamefont {C.}~\bibnamefont {Mommen}},\
  }\bibfield  {title} {\enquote {\bibinfo {title} {Complex dynamics and musical
  structure},}\ }\href@noop {} {\bibfield  {journal} {\bibinfo  {journal}
  {Interface}\ }\textbf {\bibinfo {volume} {19}},\ \bibinfo {pages} {3--14}
  (\bibinfo {year} {1990})}\BibitemShut {NoStop}%
\bibitem [{\citenamefont {Liu}\ \emph {et~al.}(2013)\citenamefont {Liu},
  \citenamefont {Wei}, \citenamefont {Zhang}, \citenamefont {Xin},\ and\
  \citenamefont {Huang}}]{liu2013}%
  \BibitemOpen
  \bibfield  {author} {\bibinfo {author} {\bibfnamefont {L.}~\bibnamefont
  {Liu}}, \bibinfo {author} {\bibfnamefont {J.}~\bibnamefont {Wei}}, \bibinfo
  {author} {\bibfnamefont {H.}~\bibnamefont {Zhang}}, \bibinfo {author}
  {\bibfnamefont {J.}~\bibnamefont {Xin}}, \ and\ \bibinfo {author}
  {\bibfnamefont {J.}~\bibnamefont {Huang}},\ }\bibfield  {title} {\enquote
  {\bibinfo {title} {A statistical physics view of pitch fluctuations in the
  classical music from bach to chopin: Evidence for scaling},}\ }\href@noop {}
  {\bibfield  {journal} {\bibinfo  {journal} {PLOS ONE}\ }\textbf {\bibinfo
  {volume} {8}},\ \bibinfo {pages} {1--6} (\bibinfo {year} {2013})}\BibitemShut
  {NoStop}%
\bibitem [{\citenamefont {Kahneman}\ \emph {et~al.}(1982)\citenamefont
  {Kahneman}, \citenamefont {Slovic}, \citenamefont {Slovic},\ and\
  \citenamefont {Tversky}}]{kahneman1982judgment}%
  \BibitemOpen
  \bibfield  {author} {\bibinfo {author} {\bibfnamefont {D.}~\bibnamefont
  {Kahneman}}, \bibinfo {author} {\bibfnamefont {S.~P.}\ \bibnamefont
  {Slovic}}, \bibinfo {author} {\bibfnamefont {P.}~\bibnamefont {Slovic}}, \
  and\ \bibinfo {author} {\bibfnamefont {A.}~\bibnamefont {Tversky}},\
  }\href@noop {} {\emph {\bibinfo {title} {Judgment under uncertainty:
  Heuristics and biases}}}\ (\bibinfo  {publisher} {Cambridge university
  press},\ \bibinfo {year} {1982})\BibitemShut {NoStop}%
\bibitem [{\citenamefont {Dayan}(1993)}]{dayan1993improving}%
  \BibitemOpen
  \bibfield  {author} {\bibinfo {author} {\bibfnamefont {P.}~\bibnamefont
  {Dayan}},\ }\bibfield  {title} {\enquote {\bibinfo {title} {Improving
  generalization for temporal difference learning: The successor
  representation},}\ }\href@noop {} {\bibfield  {journal} {\bibinfo  {journal}
  {Neural computation}\ }\textbf {\bibinfo {volume} {5}},\ \bibinfo {pages}
  {613--624} (\bibinfo {year} {1993})}\BibitemShut {NoStop}%
\bibitem [{\citenamefont {{C. W. Lynn, A. E. Kahn, N. Nyema, D. S.
  Bassett}}(2020)}]{clynn:natcom2020}%
  \BibitemOpen
  \bibfield  {author} {\bibinfo {author} {\bibnamefont {{C. W. Lynn, A. E.
  Kahn, N. Nyema, D. S. Bassett}}},\ }\bibfield  {title} {\enquote {\bibinfo
  {title} {{Abstract representations of events arise from mental errors in
  learning and memory}},}\ }\href@noop {} {\bibfield  {journal} {\bibinfo
  {journal} {Nature Communications}\ ,\ \bibinfo {pages} {2313}} (\bibinfo
  {year} {2020})}\BibitemShut {NoStop}%
\bibitem [{\citenamefont {Momennejad}\ \emph {et~al.}(2017)\citenamefont
  {Momennejad}, \citenamefont {Russek}, \citenamefont {Cheong}, \citenamefont
  {Botvinick}, \citenamefont {Daw},\ and\ \citenamefont
  {Gershman}}]{momennejad2017successor}%
  \BibitemOpen
  \bibfield  {author} {\bibinfo {author} {\bibfnamefont {I.}~\bibnamefont
  {Momennejad}}, \bibinfo {author} {\bibfnamefont {E.~M.}\ \bibnamefont
  {Russek}}, \bibinfo {author} {\bibfnamefont {J.~H.}\ \bibnamefont {Cheong}},
  \bibinfo {author} {\bibfnamefont {M.~M.}\ \bibnamefont {Botvinick}}, \bibinfo
  {author} {\bibfnamefont {N.~D.}\ \bibnamefont {Daw}}, \ and\ \bibinfo
  {author} {\bibfnamefont {S.~J.}\ \bibnamefont {Gershman}},\ }\bibfield
  {title} {\enquote {\bibinfo {title} {The successor representation in human
  reinforcement learning},}\ }\href@noop {} {\bibfield  {journal} {\bibinfo
  {journal} {Nature human behaviour}\ }\textbf {\bibinfo {volume} {1}},\
  \bibinfo {pages} {680--692} (\bibinfo {year} {2017})}\BibitemShut {NoStop}%
\bibitem [{\citenamefont {Howard}\ and\ \citenamefont
  {Kahana}(2002)}]{howard2002distributed}%
  \BibitemOpen
  \bibfield  {author} {\bibinfo {author} {\bibfnamefont {M.~W.}\ \bibnamefont
  {Howard}}\ and\ \bibinfo {author} {\bibfnamefont {M.~J.}\ \bibnamefont
  {Kahana}},\ }\bibfield  {title} {\enquote {\bibinfo {title} {A distributed
  representation of temporal context},}\ }\href@noop {} {\bibfield  {journal}
  {\bibinfo  {journal} {Journal of mathematical psychology}\ }\textbf {\bibinfo
  {volume} {46}},\ \bibinfo {pages} {269--299} (\bibinfo {year}
  {2002})}\BibitemShut {NoStop}%
\bibitem [{\citenamefont {Gershman}\ \emph {et~al.}(2012)\citenamefont
  {Gershman}, \citenamefont {Moore}, \citenamefont {Todd}, \citenamefont
  {Norman},\ and\ \citenamefont {Sederberg}}]{gershman2012successor}%
  \BibitemOpen
  \bibfield  {author} {\bibinfo {author} {\bibfnamefont {S.~J.}\ \bibnamefont
  {Gershman}}, \bibinfo {author} {\bibfnamefont {C.~D.}\ \bibnamefont {Moore}},
  \bibinfo {author} {\bibfnamefont {M.~T.}\ \bibnamefont {Todd}}, \bibinfo
  {author} {\bibfnamefont {K.~A.}\ \bibnamefont {Norman}}, \ and\ \bibinfo
  {author} {\bibfnamefont {P.~B.}\ \bibnamefont {Sederberg}},\ }\bibfield
  {title} {\enquote {\bibinfo {title} {The successor representation and
  temporal context},}\ }\href@noop {} {\bibfield  {journal} {\bibinfo
  {journal} {Neural Computation}\ }\textbf {\bibinfo {volume} {24}},\ \bibinfo
  {pages} {1553--1568} (\bibinfo {year} {2012})}\BibitemShut {NoStop}%
\bibitem [{\citenamefont {Lynn}\ and\ \citenamefont
  {Bassett}(2020)}]{clynn:pnas2020}%
  \BibitemOpen
  \bibfield  {author} {\bibinfo {author} {\bibfnamefont {C.~W.}\ \bibnamefont
  {Lynn}}\ and\ \bibinfo {author} {\bibfnamefont {D.~S.}\ \bibnamefont
  {Bassett}},\ }\bibfield  {title} {\enquote {\bibinfo {title} {{How humans
  learn and represent networks}},}\ }\href@noop {} {\bibfield  {journal}
  {\bibinfo  {journal} {Proceedings of the National Academy of Sciences}\
  }\textbf {\bibinfo {volume} {117}},\ \bibinfo {pages} {29407} (\bibinfo
  {year} {2020})}\BibitemShut {NoStop}%
\bibitem [{\citenamefont {{G. R. Jagari, P. Pedram, L.
  Hedayatifar}}(2007)}]{jafari}%
  \BibitemOpen
  \bibfield  {author} {\bibinfo {author} {\bibnamefont {{G. R. Jagari, P.
  Pedram, L. Hedayatifar}}},\ }\bibfield  {title} {\enquote {\bibinfo {title}
  {{Long-range correlation and multifractality in Bach's invention pitches}},}\
  }\href@noop {} {\bibfield  {journal} {\bibinfo  {journal} {Journal of
  Statistical Mechanics: Theory and Experiment}\ }\textbf {\bibinfo {volume}
  {04}} (\bibinfo {year} {2007})}\BibitemShut {NoStop}%
\bibitem [{\citenamefont {Moore}\ \emph {et~al.}(2018)\citenamefont {Moore},
  \citenamefont {Corrêa},\ and\ \citenamefont {Small}}]{bach_not_markov}%
  \BibitemOpen
  \bibfield  {author} {\bibinfo {author} {\bibfnamefont {J.~M.}\ \bibnamefont
  {Moore}}, \bibinfo {author} {\bibfnamefont {D.~C.}\ \bibnamefont {Corrêa}},
  \ and\ \bibinfo {author} {\bibfnamefont {M.}~\bibnamefont {Small}},\
  }\bibfield  {title} {\enquote {\bibinfo {title} {Is bach’s brain a markov
  chain? recurrence quantification to assess markov order for short, symbolic,
  musical compositions},}\ }\href@noop {} {\bibfield  {journal} {\bibinfo
  {journal} {Chaos: An Interdisciplinary Journal of Nonlinear Science}\
  }\textbf {\bibinfo {volume} {28}},\ \bibinfo {pages} {085715} (\bibinfo
  {year} {2018})}\BibitemShut {NoStop}%
\bibitem [{\citenamefont {Lerdahl}\ and\ \citenamefont
  {Jackendoff}(1996)}]{lerdahl1996generative}%
  \BibitemOpen
  \bibfield  {author} {\bibinfo {author} {\bibfnamefont {F.}~\bibnamefont
  {Lerdahl}}\ and\ \bibinfo {author} {\bibfnamefont {R.~S.}\ \bibnamefont
  {Jackendoff}},\ }\href@noop {} {\emph {\bibinfo {title} {A Generative Theory
  of Tonal Music, reissue, with a new preface}}}\ (\bibinfo  {publisher} {MIT
  press},\ \bibinfo {year} {1996})\BibitemShut {NoStop}%
\bibitem [{\citenamefont {Koelsch}\ \emph {et~al.}(2013)\citenamefont
  {Koelsch}, \citenamefont {Rohrmeier}, \citenamefont {Torrecuso},\ and\
  \citenamefont {Jentschke}}]{koelsch2013processing}%
  \BibitemOpen
  \bibfield  {author} {\bibinfo {author} {\bibfnamefont {S.}~\bibnamefont
  {Koelsch}}, \bibinfo {author} {\bibfnamefont {M.}~\bibnamefont {Rohrmeier}},
  \bibinfo {author} {\bibfnamefont {R.}~\bibnamefont {Torrecuso}}, \ and\
  \bibinfo {author} {\bibfnamefont {S.}~\bibnamefont {Jentschke}},\ }\bibfield
  {title} {\enquote {\bibinfo {title} {Processing of hierarchical syntactic
  structure in music},}\ }\href@noop {} {\bibfield  {journal} {\bibinfo
  {journal} {Proceedings of the National Academy of Sciences}\ }\textbf
  {\bibinfo {volume} {110}},\ \bibinfo {pages} {15443--15448} (\bibinfo {year}
  {2013})}\BibitemShut {NoStop}%
\bibitem [{\citenamefont {{{J. Gomez-Gardenes} and V.
  Latora}}(2008)}]{gomez2008}%
  \BibitemOpen
  \bibfield  {author} {\bibinfo {author} {\bibnamefont {{{J. Gomez-Gardenes}
  and V. Latora}}},\ }\bibfield  {title} {\enquote {\bibinfo {title} {{Entropy
  rate of diffusion processes on complex networks}},}\ }\href@noop {}
  {\bibfield  {journal} {\bibinfo  {journal} {Physical Review E}\ }\textbf
  {\bibinfo {volume} {78}},\ \bibinfo {pages} {065102} (\bibinfo {year}
  {2008})}\BibitemShut {NoStop}%
\bibitem [{\citenamefont {Albert}(2005)}]{albert2005scale}%
  \BibitemOpen
  \bibfield  {author} {\bibinfo {author} {\bibfnamefont {R.}~\bibnamefont
  {Albert}},\ }\bibfield  {title} {\enquote {\bibinfo {title} {Scale-free
  networks in cell biology},}\ }\href@noop {} {\bibfield  {journal} {\bibinfo
  {journal} {Journal of cell science}\ }\textbf {\bibinfo {volume} {118}},\
  \bibinfo {pages} {4947--4957} (\bibinfo {year} {2005})}\BibitemShut {NoStop}%
\bibitem [{\citenamefont {Lerman}\ \emph {et~al.}(2016)\citenamefont {Lerman},
  \citenamefont {Yan},\ and\ \citenamefont {Wu}}]{lerman2016majority}%
  \BibitemOpen
  \bibfield  {author} {\bibinfo {author} {\bibfnamefont {K.}~\bibnamefont
  {Lerman}}, \bibinfo {author} {\bibfnamefont {X.}~\bibnamefont {Yan}}, \ and\
  \bibinfo {author} {\bibfnamefont {X.-Z.}\ \bibnamefont {Wu}},\ }\bibfield
  {title} {\enquote {\bibinfo {title} {The majority illusion in social
  networks},}\ }\href@noop {} {\bibfield  {journal} {\bibinfo  {journal} {PloS
  one}\ }\textbf {\bibinfo {volume} {11}},\ \bibinfo {pages} {e0147617}
  (\bibinfo {year} {2016})}\BibitemShut {NoStop}%
\bibitem [{\citenamefont {Barabási}\ and\ \citenamefont
  {Albert}(1999)}]{BarabasiAlbert1999scalefree}%
  \BibitemOpen
  \bibfield  {author} {\bibinfo {author} {\bibfnamefont {A.-L.}\ \bibnamefont
  {Barabási}}\ and\ \bibinfo {author} {\bibfnamefont {R.}~\bibnamefont
  {Albert}},\ }\bibfield  {title} {\enquote {\bibinfo {title} {Emergence of
  scaling in random networks},}\ }\href@noop {} {\bibfield  {journal} {\bibinfo
   {journal} {Science}\ }\textbf {\bibinfo {volume} {286}},\ \bibinfo {pages}
  {509--512} (\bibinfo {year} {1999})}\BibitemShut {NoStop}%
\bibitem [{\citenamefont {Garvert}\ \emph {et~al.}(2017)\citenamefont
  {Garvert}, \citenamefont {Dolan},\ and\ \citenamefont
  {Behrens}}]{garvert2017map}%
  \BibitemOpen
  \bibfield  {author} {\bibinfo {author} {\bibfnamefont {M.~M.}\ \bibnamefont
  {Garvert}}, \bibinfo {author} {\bibfnamefont {R.~J.}\ \bibnamefont {Dolan}},
  \ and\ \bibinfo {author} {\bibfnamefont {T.~EJ}\ \bibnamefont {Behrens}},\
  }\bibfield  {title} {\enquote {\bibinfo {title} {A map of abstract relational
  knowledge in the human hippocampal--entorhinal cortex},}\ }\href@noop {}
  {\bibfield  {journal} {\bibinfo  {journal} {elife}\ }\textbf {\bibinfo
  {volume} {6}} (\bibinfo {year} {2017})}\BibitemShut {NoStop}%
\bibitem [{\citenamefont {Meyniel}\ \emph {et~al.}(2016)\citenamefont
  {Meyniel}, \citenamefont {Maheu},\ and\ \citenamefont
  {Dehaene}}]{meyniel2016human}%
  \BibitemOpen
  \bibfield  {author} {\bibinfo {author} {\bibfnamefont {F.}~\bibnamefont
  {Meyniel}}, \bibinfo {author} {\bibfnamefont {M.}~\bibnamefont {Maheu}}, \
  and\ \bibinfo {author} {\bibfnamefont {S.}~\bibnamefont {Dehaene}},\
  }\bibfield  {title} {\enquote {\bibinfo {title} {Human inferences about
  sequences: A minimal transition probability model},}\ }\href@noop {}
  {\bibfield  {journal} {\bibinfo  {journal} {PLoS computational biology}\
  }\textbf {\bibinfo {volume} {12}},\ \bibinfo {pages} {e1005260} (\bibinfo
  {year} {2016})}\BibitemShut {NoStop}%
\bibitem [{\citenamefont {Meyniel}\ and\ \citenamefont
  {Dehaene}(2017)}]{meyniel2017brain}%
  \BibitemOpen
  \bibfield  {author} {\bibinfo {author} {\bibfnamefont {F.}~\bibnamefont
  {Meyniel}}\ and\ \bibinfo {author} {\bibfnamefont {S.}~\bibnamefont
  {Dehaene}},\ }\bibfield  {title} {\enquote {\bibinfo {title} {Brain networks
  for confidence weighting and hierarchical inference during probabilistic
  learning},}\ }\href@noop {} {\bibfield  {journal} {\bibinfo  {journal}
  {Proceedings of the National Academy of Sciences}\ }\textbf {\bibinfo
  {volume} {114}},\ \bibinfo {pages} {E3859--E3868} (\bibinfo {year}
  {2017})}\BibitemShut {NoStop}%
\bibitem [{\citenamefont {Newport}\ and\ \citenamefont
  {Aslin}(2004)}]{newport2004learning}%
  \BibitemOpen
  \bibfield  {author} {\bibinfo {author} {\bibfnamefont {E.~L.}\ \bibnamefont
  {Newport}}\ and\ \bibinfo {author} {\bibfnamefont {R.~N.}\ \bibnamefont
  {Aslin}},\ }\bibfield  {title} {\enquote {\bibinfo {title} {Learning at a
  distance i. statistical learning of non-adjacent dependencies},}\ }\href@noop
  {} {\bibfield  {journal} {\bibinfo  {journal} {Cognitive psychology}\
  }\textbf {\bibinfo {volume} {48}},\ \bibinfo {pages} {127--162} (\bibinfo
  {year} {2004})}\BibitemShut {NoStop}%
\bibitem [{\citenamefont {{A. E. Kahn, E. A. Karuza, J. M. Vettel, D. S.
  Bassett}}(2018)}]{kahn2018}%
  \BibitemOpen
  \bibfield  {author} {\bibinfo {author} {\bibnamefont {{A. E. Kahn, E. A.
  Karuza, J. M. Vettel, D. S. Bassett}}},\ }\bibfield  {title} {\enquote
  {\bibinfo {title} {{Network constraints on learnability of probabilistic
  motor sequences}},}\ }\href@noop {} {\bibfield  {journal} {\bibinfo
  {journal} {Nature Human Behavior}\ ,\ \bibinfo {pages} {936--947}} (\bibinfo
  {year} {2018})}\BibitemShut {NoStop}%
\bibitem [{\citenamefont {Saffran}\ \emph {et~al.}(1999)\citenamefont
  {Saffran}, \citenamefont {Johnson}, \citenamefont {Aslin},\ and\
  \citenamefont {Newport}}]{saffran1999}%
  \BibitemOpen
  \bibfield  {author} {\bibinfo {author} {\bibfnamefont {J.~R.}\ \bibnamefont
  {Saffran}}, \bibinfo {author} {\bibfnamefont {E.~K.}\ \bibnamefont
  {Johnson}}, \bibinfo {author} {\bibfnamefont {R.~N.}\ \bibnamefont {Aslin}},
  \ and\ \bibinfo {author} {\bibfnamefont {E.~L.}\ \bibnamefont {Newport}},\
  }\bibfield  {title} {\enquote {\bibinfo {title} {Statistical learning of tone
  sequences by human infants and adults},}\ }\href@noop {} {\bibfield
  {journal} {\bibinfo  {journal} {Cognition}\ }\textbf {\bibinfo {volume} {70}}
  (\bibinfo {year} {1999})}\BibitemShut {NoStop}%
\bibitem [{\citenamefont {Morgan}\ \emph {et~al.}(2019)\citenamefont {Morgan},
  \citenamefont {Fogel}, \citenamefont {Nair},\ and\ \citenamefont
  {Patel}}]{morgan2019}%
  \BibitemOpen
  \bibfield  {author} {\bibinfo {author} {\bibfnamefont {E.}~\bibnamefont
  {Morgan}}, \bibinfo {author} {\bibfnamefont {A.}~\bibnamefont {Fogel}},
  \bibinfo {author} {\bibfnamefont {A.}~\bibnamefont {Nair}}, \ and\ \bibinfo
  {author} {\bibfnamefont {A.~D.}\ \bibnamefont {Patel}},\ }\bibfield  {title}
  {\enquote {\bibinfo {title} {Statistical learning and gestalt-like principles
  predict melodic expectations},}\ }\href@noop {} {\bibfield  {journal}
  {\bibinfo  {journal} {Cognition}\ }\textbf {\bibinfo {volume} {189}}
  (\bibinfo {year} {2019})}\BibitemShut {NoStop}%
\bibitem [{\citenamefont {Cheung}\ \emph {et~al.}(2020)\citenamefont {Cheung},
  \citenamefont {Harrison}, \citenamefont {Koelsch}, \citenamefont {Pearce},
  \citenamefont {Friederici},\ and\ \citenamefont {Meyer}}]{cheung_2020}%
  \BibitemOpen
  \bibfield  {author} {\bibinfo {author} {\bibfnamefont {V.~K.}\ \bibnamefont
  {Cheung}}, \bibinfo {author} {\bibfnamefont {P.~M.~C.}\ \bibnamefont
  {Harrison}}, \bibinfo {author} {\bibfnamefont {S.}~\bibnamefont {Koelsch}},
  \bibinfo {author} {\bibfnamefont {M.~T.}\ \bibnamefont {Pearce}}, \bibinfo
  {author} {\bibfnamefont {A.~D.}\ \bibnamefont {Friederici}}, \ and\ \bibinfo
  {author} {\bibfnamefont {L.}~\bibnamefont {Meyer}},\ }\href@noop {} {\enquote
  {\bibinfo {title} {Distinct roles of cognitive and sensory information in
  musical expectancy},}\ } (\bibinfo {year} {2020})\BibitemShut {NoStop}%
\bibitem [{\citenamefont {Collins}\ \emph {et~al.}(2014)\citenamefont
  {Collins}, \citenamefont {Tillmann}, \citenamefont {Barrett}, \citenamefont
  {Delb{\'e}},\ and\ \citenamefont {Janata}}]{collins2014combined}%
  \BibitemOpen
  \bibfield  {author} {\bibinfo {author} {\bibfnamefont {T.}~\bibnamefont
  {Collins}}, \bibinfo {author} {\bibfnamefont {Barbara}\ \bibnamefont
  {Tillmann}}, \bibinfo {author} {\bibfnamefont {Frederick~S}\ \bibnamefont
  {Barrett}}, \bibinfo {author} {\bibfnamefont {Charles}\ \bibnamefont
  {Delb{\'e}}}, \ and\ \bibinfo {author} {\bibfnamefont {Petr}\ \bibnamefont
  {Janata}},\ }\bibfield  {title} {\enquote {\bibinfo {title} {A combined model
  of sensory and cognitive representations underlying tonal expectations in
  music: from audio signals to behavior.}}\ }\href@noop {} {\bibfield
  {journal} {\bibinfo  {journal} {Psychological review}\ }\textbf {\bibinfo
  {volume} {121}} (\bibinfo {year} {2014})}\BibitemShut {NoStop}%
\bibitem [{\citenamefont {Watts}\ and\ \citenamefont
  {Strogatz}(1998{\natexlab{a}})}]{smallworld}%
  \BibitemOpen
  \bibfield  {author} {\bibinfo {author} {\bibfnamefont {D.~J.}\ \bibnamefont
  {Watts}}\ and\ \bibinfo {author} {\bibfnamefont {S.~H.}\ \bibnamefont
  {Strogatz}},\ }\bibfield  {title} {\enquote {\bibinfo {title} {Collective
  dynamics of `small-world' networks},}\ }\href@noop {} {\bibfield  {journal}
  {\bibinfo  {journal} {Nature}\ }\textbf {\bibinfo {volume} {393}},\ \bibinfo
  {pages} {440--442} (\bibinfo {year} {1998}{\natexlab{a}})}\BibitemShut
  {NoStop}%
\bibitem [{\citenamefont {Szab{\'o}}\ \emph {et~al.}(2004)\citenamefont
  {Szab{\'o}}, \citenamefont {Alava},\ and\ \citenamefont
  {Kert{\'e}sz}}]{Szab2004Clustering}%
  \BibitemOpen
  \bibfield  {author} {\bibinfo {author} {\bibfnamefont {G.}~\bibnamefont
  {Szab{\'o}}}, \bibinfo {author} {\bibfnamefont {M.~J.}\ \bibnamefont
  {Alava}}, \ and\ \bibinfo {author} {\bibfnamefont {J.}~\bibnamefont
  {Kert{\'e}sz}},\ }\bibfield  {title} {\enquote {\bibinfo {title} {Clustering
  in complex networks},}\ }\href@noop {} {\bibfield  {journal} {\bibinfo
  {journal} {Lecture Notes in Physics}\ }\textbf {\bibinfo {volume} {650}},\
  \bibinfo {pages} {139--162} (\bibinfo {year} {2004})}\BibitemShut {NoStop}%
\bibitem [{\citenamefont {Saffran}\ \emph {et~al.}(1996)\citenamefont
  {Saffran}, \citenamefont {Aslin},\ and\ \citenamefont
  {Newport}}]{saffran1996statistical}%
  \BibitemOpen
  \bibfield  {author} {\bibinfo {author} {\bibfnamefont {Jenny~R}\ \bibnamefont
  {Saffran}}, \bibinfo {author} {\bibfnamefont {Richard~N}\ \bibnamefont
  {Aslin}}, \ and\ \bibinfo {author} {\bibfnamefont {Elissa~L}\ \bibnamefont
  {Newport}},\ }\bibfield  {title} {\enquote {\bibinfo {title} {Statistical
  learning by 8-month-old infants},}\ }\href@noop {} {\bibfield  {journal}
  {\bibinfo  {journal} {Science}\ }\textbf {\bibinfo {volume} {274}} (\bibinfo
  {year} {1996})}\BibitemShut {NoStop}%
\bibitem [{\citenamefont {Fiser}\ and\ \citenamefont
  {Aslin}(2002)}]{fiser2002statistical}%
  \BibitemOpen
  \bibfield  {author} {\bibinfo {author} {\bibfnamefont {J{\'o}zsef}\
  \bibnamefont {Fiser}}\ and\ \bibinfo {author} {\bibfnamefont {Richard~N}\
  \bibnamefont {Aslin}},\ }\bibfield  {title} {\enquote {\bibinfo {title}
  {Statistical learning of higher-order temporal structure from visual shape
  sequences.}}\ }\href@noop {} {\bibfield  {journal} {\bibinfo  {journal}
  {Journal of Experimental Psychology: Learning, Memory, and Cognition}\
  }\textbf {\bibinfo {volume} {28}} (\bibinfo {year} {2002})}\BibitemShut
  {NoStop}%
\bibitem [{\citenamefont {Romberg}\ and\ \citenamefont
  {Saffran}(2010)}]{romberg2010statistical}%
  \BibitemOpen
  \bibfield  {author} {\bibinfo {author} {\bibfnamefont {Alexa~R}\ \bibnamefont
  {Romberg}}\ and\ \bibinfo {author} {\bibfnamefont {Jenny~R}\ \bibnamefont
  {Saffran}},\ }\bibfield  {title} {\enquote {\bibinfo {title} {Statistical
  learning and language acquisition},}\ }\href@noop {} {\bibfield  {journal}
  {\bibinfo  {journal} {Wiley Interdisciplinary Reviews: Cognitive Science}\
  }\textbf {\bibinfo {volume} {1}} (\bibinfo {year} {2010})}\BibitemShut
  {NoStop}%
\bibitem [{\citenamefont {Watts}\ and\ \citenamefont
  {Strogatz}(1998{\natexlab{b}})}]{watts1998collective}%
  \BibitemOpen
  \bibfield  {author} {\bibinfo {author} {\bibfnamefont {Duncan~J}\
  \bibnamefont {Watts}}\ and\ \bibinfo {author} {\bibfnamefont {Steven~H}\
  \bibnamefont {Strogatz}},\ }\bibfield  {title} {\enquote {\bibinfo {title}
  {Collective dynamics of ‘small-world’networks},}\ }\href@noop {}
  {\bibfield  {journal} {\bibinfo  {journal} {nature}\ } (\bibinfo {year}
  {1998}{\natexlab{b}})}\BibitemShut {NoStop}%
\bibitem [{\citenamefont {Cancho}\ and\ \citenamefont
  {Sol{\'e}}(2001)}]{cancho2001small}%
  \BibitemOpen
  \bibfield  {author} {\bibinfo {author} {\bibfnamefont {Ramon Ferrer~I}\
  \bibnamefont {Cancho}}\ and\ \bibinfo {author} {\bibfnamefont {Richard~V}\
  \bibnamefont {Sol{\'e}}},\ }\bibfield  {title} {\enquote {\bibinfo {title}
  {The small world of human language},}\ }\href@noop {} {\bibfield  {journal}
  {\bibinfo  {journal} {Proceedings of the Royal Society of London. Series B:
  Biological Sciences}\ } (\bibinfo {year} {2001})}\BibitemShut {NoStop}%
\bibitem [{\citenamefont {Zivic}\ \emph {et~al.}(2013)\citenamefont {Zivic},
  \citenamefont {Shifres},\ and\ \citenamefont {Cecchi}}]{zivic2013}%
  \BibitemOpen
  \bibfield  {author} {\bibinfo {author} {\bibfnamefont {P.~H.~R.}\
  \bibnamefont {Zivic}}, \bibinfo {author} {\bibfnamefont {F.}~\bibnamefont
  {Shifres}}, \ and\ \bibinfo {author} {\bibfnamefont {G.~A.}\ \bibnamefont
  {Cecchi}},\ }\bibfield  {title} {\enquote {\bibinfo {title} {Perceptual basis
  of evolving western musical styles},}\ }\href@noop {} {\bibfield  {journal}
  {\bibinfo  {journal} {Proceedings of the National Academy of Sciences}\
  }\textbf {\bibinfo {volume} {110}},\ \bibinfo {pages} {10034--10038}
  (\bibinfo {year} {2013})}\BibitemShut {NoStop}%
\bibitem [{\citenamefont {Pérez-Sancho}\ \emph {et~al.}(2009)\citenamefont
  {Pérez-Sancho}, \citenamefont {Rizo},\ and\ \citenamefont
  {Iñesta}}]{carlos2009}%
  \BibitemOpen
  \bibfield  {author} {\bibinfo {author} {\bibfnamefont {C.}~\bibnamefont
  {Pérez-Sancho}}, \bibinfo {author} {\bibfnamefont {D.}~\bibnamefont {Rizo}},
  \ and\ \bibinfo {author} {\bibfnamefont {J.~M.}\ \bibnamefont {Iñesta}},\
  }\bibfield  {title} {\enquote {\bibinfo {title} {Genre classification using
  chords and stochastic language models},}\ }\href@noop {} {\bibfield
  {journal} {\bibinfo  {journal} {Connection Science}\ }\textbf {\bibinfo
  {volume} {21}},\ \bibinfo {pages} {145--159} (\bibinfo {year}
  {2009})}\BibitemShut {NoStop}%
\bibitem [{\citenamefont {Simonton}(1984)}]{simonton1984}%
  \BibitemOpen
  \bibfield  {author} {\bibinfo {author} {\bibfnamefont {D.~K.}\ \bibnamefont
  {Simonton}},\ }\bibfield  {title} {\enquote {\bibinfo {title} {Melodic
  structure and note transition probabilities: A content analysis of 15,618
  classical themes},}\ }\href@noop {} {\bibfield  {journal} {\bibinfo
  {journal} {Psychology of Music}\ }\textbf {\bibinfo {volume} {12}},\ \bibinfo
  {pages} {3--16} (\bibinfo {year} {1984})}\BibitemShut {NoStop}%
\bibitem [{\citenamefont {Thickstun}\ \emph {et~al.}(2016)\citenamefont
  {Thickstun}, \citenamefont {Harchaoui},\ and\ \citenamefont
  {Kakade}}]{thickstun2016learning}%
  \BibitemOpen
  \bibfield  {author} {\bibinfo {author} {\bibfnamefont {J.}~\bibnamefont
  {Thickstun}}, \bibinfo {author} {\bibfnamefont {Z.}~\bibnamefont
  {Harchaoui}}, \ and\ \bibinfo {author} {\bibfnamefont {S.}~\bibnamefont
  {Kakade}},\ }\bibfield  {title} {\enquote {\bibinfo {title} {Learning
  features of music from scratch},}\ }\href@noop {} {\bibfield  {journal}
  {\bibinfo  {journal} {arXiv preprint arXiv:1611.09827}\ } (\bibinfo {year}
  {2016})}\BibitemShut {NoStop}%
\bibitem [{\citenamefont {Liu}\ \emph {et~al.}(2014)\citenamefont {Liu},
  \citenamefont {Ramakrishnan} \emph {et~al.}}]{liu2014bach}%
  \BibitemOpen
  \bibfield  {author} {\bibinfo {author} {\bibfnamefont {I}~\bibnamefont
  {Liu}}, \bibinfo {author} {\bibfnamefont {Bhiksha}\ \bibnamefont
  {Ramakrishnan}},  \emph {et~al.},\ }\bibfield  {title} {\enquote {\bibinfo
  {title} {Bach in 2014: Music composition with recurrent neural network},}\
  }\href@noop {} {\bibfield  {journal} {\bibinfo  {journal} {arXiv preprint
  arXiv:1412.3191}\ } (\bibinfo {year} {2014})}\BibitemShut {NoStop}%
\bibitem [{\citenamefont {Dowling}\ and\ \citenamefont
  {Bartlett}(1981)}]{dowling1981importance}%
  \BibitemOpen
  \bibfield  {author} {\bibinfo {author} {\bibfnamefont {W~Jay}\ \bibnamefont
  {Dowling}}\ and\ \bibinfo {author} {\bibfnamefont {James~C}\ \bibnamefont
  {Bartlett}},\ }\bibfield  {title} {\enquote {\bibinfo {title} {The importance
  of interval information in long-term memory for melodies.}}\ }\href@noop {}
  {\bibfield  {journal} {\bibinfo  {journal} {Psychomusicology: A Journal of
  Research in Music Cognition}\ }\textbf {\bibinfo {volume} {1}} (\bibinfo
  {year} {1981})}\BibitemShut {NoStop}%
\bibitem [{\citenamefont {Parncutt}\ and\ \citenamefont
  {Parncutt}(1989)}]{parncutt1989psychoacoustics}%
  \BibitemOpen
  \bibfield  {author} {\bibinfo {author} {\bibfnamefont {Richard}\ \bibnamefont
  {Parncutt}}\ and\ \bibinfo {author} {\bibfnamefont {Richard}\ \bibnamefont
  {Parncutt}},\ }\bibfield  {title} {\enquote {\bibinfo {title}
  {Psychoacoustics},}\ }\href@noop {} {\bibfield  {journal} {\bibinfo
  {journal} {Harmony: a psychoacoustical approach}\ } (\bibinfo {year}
  {1989})}\BibitemShut {NoStop}%
\bibitem [{\citenamefont {Xu}\ \emph {et~al.}(2016)\citenamefont {Xu},
  \citenamefont {Wickramarathne},\ and\ \citenamefont
  {Chawla}}]{xu2016representing}%
  \BibitemOpen
  \bibfield  {author} {\bibinfo {author} {\bibfnamefont {Jian}\ \bibnamefont
  {Xu}}, \bibinfo {author} {\bibfnamefont {Thanuka~L}\ \bibnamefont
  {Wickramarathne}}, \ and\ \bibinfo {author} {\bibfnamefont {Nitesh~V}\
  \bibnamefont {Chawla}},\ }\bibfield  {title} {\enquote {\bibinfo {title}
  {Representing higher-order dependencies in networks},}\ }\href@noop {}
  {\bibfield  {journal} {\bibinfo  {journal} {Science advances}\ }\textbf
  {\bibinfo {volume} {2}} (\bibinfo {year} {2016})}\BibitemShut {NoStop}%
\bibitem [{\citenamefont {Lambiotte}\ \emph {et~al.}(2019)\citenamefont
  {Lambiotte}, \citenamefont {Rosvall},\ and\ \citenamefont
  {Scholtes}}]{lambiotte2019networks}%
  \BibitemOpen
  \bibfield  {author} {\bibinfo {author} {\bibfnamefont {Renaud}\ \bibnamefont
  {Lambiotte}}, \bibinfo {author} {\bibfnamefont {Martin}\ \bibnamefont
  {Rosvall}}, \ and\ \bibinfo {author} {\bibfnamefont {Ingo}\ \bibnamefont
  {Scholtes}},\ }\bibfield  {title} {\enquote {\bibinfo {title} {From networks
  to optimal higher-order models of complex systems},}\ }\href@noop {}
  {\bibfield  {journal} {\bibinfo  {journal} {Nature physics}\ }\textbf
  {\bibinfo {volume} {15}} (\bibinfo {year} {2019})}\BibitemShut {NoStop}%
\bibitem [{\citenamefont {Yin}\ \emph {et~al.}(2018)\citenamefont {Yin},
  \citenamefont {Benson},\ and\ \citenamefont {Leskovec}}]{yin2018higher}%
  \BibitemOpen
  \bibfield  {author} {\bibinfo {author} {\bibfnamefont {Hao}\ \bibnamefont
  {Yin}}, \bibinfo {author} {\bibfnamefont {Austin~R}\ \bibnamefont {Benson}},
  \ and\ \bibinfo {author} {\bibfnamefont {Jure}\ \bibnamefont {Leskovec}},\
  }\bibfield  {title} {\enquote {\bibinfo {title} {Higher-order clustering in
  networks},}\ }\href@noop {} {\bibfield  {journal} {\bibinfo  {journal}
  {Physical Review E}\ }\textbf {\bibinfo {volume} {97}} (\bibinfo {year}
  {2018})}\BibitemShut {NoStop}%
\bibitem [{\citenamefont {Stiller}\ \emph {et~al.}(2003)\citenamefont
  {Stiller}, \citenamefont {Nettle},\ and\ \citenamefont
  {Dunbar}}]{stiller2003small}%
  \BibitemOpen
  \bibfield  {author} {\bibinfo {author} {\bibfnamefont {J.}~\bibnamefont
  {Stiller}}, \bibinfo {author} {\bibfnamefont {D.}~\bibnamefont {Nettle}}, \
  and\ \bibinfo {author} {\bibfnamefont {R.~IM}\ \bibnamefont {Dunbar}},\
  }\bibfield  {title} {\enquote {\bibinfo {title} {The small world of
  shakespeare’s plays},}\ }\href@noop {} {\bibfield  {journal} {\bibinfo
  {journal} {Human Nature}\ }\textbf {\bibinfo {volume} {14}},\ \bibinfo
  {pages} {397--408} (\bibinfo {year} {2003})}\BibitemShut {NoStop}%
\bibitem [{\citenamefont {Choi}\ and\ \citenamefont
  {Kim}(2007)}]{choi2007directed}%
  \BibitemOpen
  \bibfield  {author} {\bibinfo {author} {\bibfnamefont {Y.-M.}\ \bibnamefont
  {Choi}}\ and\ \bibinfo {author} {\bibfnamefont {H.-J.}\ \bibnamefont {Kim}},\
  }\bibfield  {title} {\enquote {\bibinfo {title} {A directed network of greek
  and roman mythology},}\ }\href@noop {} {\bibfield  {journal} {\bibinfo
  {journal} {Physica A: statistical Mechanics and its Applications}\ }\textbf
  {\bibinfo {volume} {382}},\ \bibinfo {pages} {665--671} (\bibinfo {year}
  {2007})}\BibitemShut {NoStop}%
\bibitem [{bc()}]{bc}%
  \BibitemOpen
  \href@noop {} {\enquote {\bibinfo {title} {Bach central},}\ }\bibinfo
  {howpublished}
  {\url{http://www.bachcentral.com/midiindexcomplete.html}}\BibitemShut
  {NoStop}%
\bibitem [{ks()}]{ks}%
  \BibitemOpen
  \href@noop {} {\enquote {\bibinfo {title} {Kern scores},}\ }\bibinfo
  {howpublished}
  {\url{http://kern.humdrum.org/search?s=t&keyword=Bach+Johann}}\BibitemShut
  {NoStop}%
\bibitem [{bcw()}]{bcw}%
  \BibitemOpen
  \href@noop {} {\enquote {\bibinfo {title} {Bach cantatas website},}\
  }\bibinfo {howpublished}
  {\url{https://www.bach-cantatas.com/Mus/BWV1-Mus.htm}}\BibitemShut {NoStop}%
\bibitem [{sm()}]{sm}%
  \BibitemOpen
  \href@noop {} {\enquote {\bibinfo {title} {Suzu midi},}\ }\bibinfo
  {howpublished} {\url{http://www.suzumidi.com/eng/bach4.htm}}\BibitemShut
  {NoStop}%
\bibitem [{mid()}]{midi}%
  \BibitemOpen
  \href@noop {} {\enquote {\bibinfo {title} {Midi file tools for matlab by ken
  schutte},}\ }\bibinfo {howpublished}
  {\url{https://github.com/kts/matlab-midi}}\BibitemShut {NoStop}%
\bibitem [{SK_()}]{SK_code}%
  \BibitemOpen
  \href@noop {} {\enquote {\bibinfo {title} {Information content of note
  transitions in the music of j. s. bach.}}\ }\bibinfo {howpublished}
  {\url{https://github.com/SumanSKulkarni/Music_Networks}}\BibitemShut
  {NoStop}%
\bibitem [{\citenamefont {Newman}\ and\ \citenamefont
  {Girvan}(2004)}]{newman2004finding}%
  \BibitemOpen
  \bibfield  {author} {\bibinfo {author} {\bibfnamefont {M.~E.~J.}\
  \bibnamefont {Newman}}\ and\ \bibinfo {author} {\bibfnamefont
  {M.}~\bibnamefont {Girvan}},\ }\bibfield  {title} {\enquote {\bibinfo {title}
  {Finding and evaluating community structure in networks},}\ }\href@noop {}
  {\bibfield  {journal} {\bibinfo  {journal} {Physical review E}\ }\textbf
  {\bibinfo {volume} {69}},\ \bibinfo {pages} {026113} (\bibinfo {year}
  {2004})}\BibitemShut {NoStop}%
\bibitem [{\citenamefont {V{\'a}{\v{s}}a}\ and\ \citenamefont
  {Mi{\v{s}}i{\'c}}(2022)}]{vavsa2022null}%
  \BibitemOpen
  \bibfield  {author} {\bibinfo {author} {\bibfnamefont {F.}~\bibnamefont
  {V{\'a}{\v{s}}a}}\ and\ \bibinfo {author} {\bibfnamefont {B.}~\bibnamefont
  {Mi{\v{s}}i{\'c}}},\ }\bibfield  {title} {\enquote {\bibinfo {title} {Null
  models in network neuroscience},}\ }\href@noop {} {\bibfield  {journal}
  {\bibinfo  {journal} {Nature Reviews Neuroscience}\ ,\ \bibinfo {pages}
  {1--12}} (\bibinfo {year} {2022})}\BibitemShut {NoStop}%
\bibitem [{\citenamefont {{G. Fagiolo}}(2007)}]{fagiolo2007}%
  \BibitemOpen
  \bibfield  {author} {\bibinfo {author} {\bibnamefont {{G. Fagiolo}}},\
  }\bibfield  {title} {\enquote {\bibinfo {title} {Clustering in complex
  directed networks},}\ }\href@noop {} {\bibfield  {journal} {\bibinfo
  {journal} {Physical Review E}\ }\textbf {\bibinfo {volume} {76}},\ \bibinfo
  {pages} {026107} (\bibinfo {year} {2007})}\BibitemShut {NoStop}%
\bibitem [{\citenamefont {Mitchell}\ \emph {et~al.}(2013)\citenamefont
  {Mitchell}, \citenamefont {Lange},\ and\ \citenamefont
  {Brus}}]{mitchell2013gendered}%
  \BibitemOpen
  \bibfield  {author} {\bibinfo {author} {\bibfnamefont {S.~M.}\ \bibnamefont
  {Mitchell}}, \bibinfo {author} {\bibfnamefont {S.}~\bibnamefont {Lange}}, \
  and\ \bibinfo {author} {\bibfnamefont {H.}~\bibnamefont {Brus}},\ }\bibfield
  {title} {\enquote {\bibinfo {title} {Gendered citation patterns in
  international relations journals},}\ }\href@noop {} {\bibfield  {journal}
  {\bibinfo  {journal} {International Studies Perspectives}\ }\textbf {\bibinfo
  {volume} {14}},\ \bibinfo {pages} {485--492} (\bibinfo {year}
  {2013})}\BibitemShut {NoStop}%
\bibitem [{\citenamefont {Dion}\ \emph {et~al.}(2018)\citenamefont {Dion},
  \citenamefont {Sumner},\ and\ \citenamefont {Mitchell}}]{dion2018gendered}%
  \BibitemOpen
  \bibfield  {author} {\bibinfo {author} {\bibfnamefont {M.~L.}\ \bibnamefont
  {Dion}}, \bibinfo {author} {\bibfnamefont {J.~L.}\ \bibnamefont {Sumner}}, \
  and\ \bibinfo {author} {\bibfnamefont {S.}~\bibnamefont {Mitchell}},\
  }\bibfield  {title} {\enquote {\bibinfo {title} {Gendered citation patterns
  across political science and social science methodology fields},}\
  }\href@noop {} {\bibfield  {journal} {\bibinfo  {journal} {Political
  Analysis}\ }\textbf {\bibinfo {volume} {26}},\ \bibinfo {pages} {312--327}
  (\bibinfo {year} {2018})}\BibitemShut {NoStop}%
\bibitem [{\citenamefont {Caplar}\ \emph {et~al.}(2017)\citenamefont {Caplar},
  \citenamefont {Tacchella},\ and\ \citenamefont
  {Birrer}}]{caplar2017quantitative}%
  \BibitemOpen
  \bibfield  {author} {\bibinfo {author} {\bibfnamefont {N.}~\bibnamefont
  {Caplar}}, \bibinfo {author} {\bibfnamefont {S.}~\bibnamefont {Tacchella}}, \
  and\ \bibinfo {author} {\bibfnamefont {S.}~\bibnamefont {Birrer}},\
  }\bibfield  {title} {\enquote {\bibinfo {title} {Quantitative evaluation of
  gender bias in astronomical publications from citation counts},}\ }\href@noop
  {} {\bibfield  {journal} {\bibinfo  {journal} {Nature Astronomy}\ }\textbf
  {\bibinfo {volume} {1}},\ \bibinfo {pages} {0141} (\bibinfo {year}
  {2017})}\BibitemShut {NoStop}%
\bibitem [{\citenamefont {Maliniak}\ \emph {et~al.}(2013)\citenamefont
  {Maliniak}, \citenamefont {Powers},\ and\ \citenamefont
  {Walter}}]{maliniak2013gender}%
  \BibitemOpen
  \bibfield  {author} {\bibinfo {author} {\bibfnamefont {D.}~\bibnamefont
  {Maliniak}}, \bibinfo {author} {\bibfnamefont {R.}~\bibnamefont {Powers}}, \
  and\ \bibinfo {author} {\bibfnamefont {B.~F.}\ \bibnamefont {Walter}},\
  }\bibfield  {title} {\enquote {\bibinfo {title} {The gender citation gap in
  international relations},}\ }\href@noop {} {\bibfield  {journal} {\bibinfo
  {journal} {International Organization}\ }\textbf {\bibinfo {volume} {67}},\
  \bibinfo {pages} {889--922} (\bibinfo {year} {2013})}\BibitemShut {NoStop}%
\bibitem [{\citenamefont {Dworkin}\ \emph {et~al.}(2020)\citenamefont
  {Dworkin}, \citenamefont {Linn}, \citenamefont {Teich}, \citenamefont {Zurn},
  \citenamefont {Shinohara},\ and\ \citenamefont
  {Bassett}}]{Dworkin2020.01.03.894378}%
  \BibitemOpen
  \bibfield  {author} {\bibinfo {author} {\bibfnamefont {J.~D.}\ \bibnamefont
  {Dworkin}}, \bibinfo {author} {\bibfnamefont {K.~A.}\ \bibnamefont {Linn}},
  \bibinfo {author} {\bibfnamefont {E.~G.}\ \bibnamefont {Teich}}, \bibinfo
  {author} {\bibfnamefont {P.}~\bibnamefont {Zurn}}, \bibinfo {author}
  {\bibfnamefont {R.~T.}\ \bibnamefont {Shinohara}}, \ and\ \bibinfo {author}
  {\bibfnamefont {D.~S.}\ \bibnamefont {Bassett}},\ }\bibfield  {title}
  {\enquote {\bibinfo {title} {The extent and drivers of gender imbalance in
  neuroscience reference lists},}\ }\href@noop {} {\bibfield  {journal}
  {\bibinfo  {journal} {bioRxiv}\ } (\bibinfo {year} {2020})}\BibitemShut
  {NoStop}%
\bibitem [{\citenamefont {Bertolero}\ \emph {et~al.}(2020)\citenamefont
  {Bertolero}, \citenamefont {Dworkin}, \citenamefont {David}, \citenamefont
  {Lloreda}, \citenamefont {Srivastava}, \citenamefont {Stiso}, \citenamefont
  {Zhou}, \citenamefont {Dzirasa}, \citenamefont {Fair}, \citenamefont
  {Kaczkurkin}, \citenamefont {Marlin}, \citenamefont {Shohamy}, \citenamefont
  {Uddin}, \citenamefont {Zurn},\ and\ \citenamefont
  {Bassett}}]{bertolero2021racial}%
  \BibitemOpen
  \bibfield  {author} {\bibinfo {author} {\bibfnamefont {M.~A.}\ \bibnamefont
  {Bertolero}}, \bibinfo {author} {\bibfnamefont {J.~D.}\ \bibnamefont
  {Dworkin}}, \bibinfo {author} {\bibfnamefont {S.~U.}\ \bibnamefont {David}},
  \bibinfo {author} {\bibfnamefont {C.~L.}\ \bibnamefont {Lloreda}}, \bibinfo
  {author} {\bibfnamefont {P.}~\bibnamefont {Srivastava}}, \bibinfo {author}
  {\bibfnamefont {J.}~\bibnamefont {Stiso}}, \bibinfo {author} {\bibfnamefont
  {D.}~\bibnamefont {Zhou}}, \bibinfo {author} {\bibfnamefont {K.}~\bibnamefont
  {Dzirasa}}, \bibinfo {author} {\bibfnamefont {D.~A.}\ \bibnamefont {Fair}},
  \bibinfo {author} {\bibfnamefont {A.~N.}\ \bibnamefont {Kaczkurkin}},
  \bibinfo {author} {\bibfnamefont {B.~J.}\ \bibnamefont {Marlin}}, \bibinfo
  {author} {\bibfnamefont {D.}~\bibnamefont {Shohamy}}, \bibinfo {author}
  {\bibfnamefont {L.~Q.}\ \bibnamefont {Uddin}}, \bibinfo {author}
  {\bibfnamefont {P.}~\bibnamefont {Zurn}}, \ and\ \bibinfo {author}
  {\bibfnamefont {D.~S.}\ \bibnamefont {Bassett}},\ }\bibfield  {title}
  {\enquote {\bibinfo {title} {Racial and ethnic imbalance in neuroscience
  reference lists and intersections with gender},}\ }\href@noop {} {\bibfield
  {journal} {\bibinfo  {journal} {bioRxiv}\ } (\bibinfo {year}
  {2020})}\BibitemShut {NoStop}%
\bibitem [{\citenamefont {Wang}\ \emph {et~al.}(2021)\citenamefont {Wang},
  \citenamefont {Dworkin}, \citenamefont {Zhou}, \citenamefont {Stiso},
  \citenamefont {Falk}, \citenamefont {Bassett}, \citenamefont {Zurn},\ and\
  \citenamefont {Lydon-Staley}}]{wang2021gendered}%
  \BibitemOpen
  \bibfield  {author} {\bibinfo {author} {\bibfnamefont {X.}~\bibnamefont
  {Wang}}, \bibinfo {author} {\bibfnamefont {J.~D.}\ \bibnamefont {Dworkin}},
  \bibinfo {author} {\bibfnamefont {D.}~\bibnamefont {Zhou}}, \bibinfo {author}
  {\bibfnamefont {J.}~\bibnamefont {Stiso}}, \bibinfo {author} {\bibfnamefont
  {E.~B.}\ \bibnamefont {Falk}}, \bibinfo {author} {\bibfnamefont {D.~S.}\
  \bibnamefont {Bassett}}, \bibinfo {author} {\bibfnamefont {P.}~\bibnamefont
  {Zurn}}, \ and\ \bibinfo {author} {\bibfnamefont {D.~M.}\ \bibnamefont
  {Lydon-Staley}},\ }\bibfield  {title} {\enquote {\bibinfo {title} {Gendered
  citation practices in the field of communication},}\ }\href@noop {}
  {\bibfield  {journal} {\bibinfo  {journal} {Annals of the International
  Communication Association}\ } (\bibinfo {year} {2021})}\BibitemShut {NoStop}%
\bibitem [{\citenamefont {Chatterjee}\ and\ \citenamefont
  {Werner}(2021)}]{chatterjee2021gender}%
  \BibitemOpen
  \bibfield  {author} {\bibinfo {author} {\bibfnamefont {P.}~\bibnamefont
  {Chatterjee}}\ and\ \bibinfo {author} {\bibfnamefont {R.~M.}\ \bibnamefont
  {Werner}},\ }\bibfield  {title} {\enquote {\bibinfo {title} {Gender disparity
  in citations in high-impact journal articles},}\ }\href@noop {} {\bibfield
  {journal} {\bibinfo  {journal} {JAMA Netw Open}\ }\textbf {\bibinfo {volume}
  {4}},\ \bibinfo {pages} {e2114509} (\bibinfo {year} {2021})}\BibitemShut
  {NoStop}%
\bibitem [{\citenamefont {Fulvio}\ \emph {et~al.}(2021)\citenamefont {Fulvio},
  \citenamefont {Akinnola},\ and\ \citenamefont
  {Postle}}]{fulvio2021imbalance}%
  \BibitemOpen
  \bibfield  {author} {\bibinfo {author} {\bibfnamefont {J.~M.}\ \bibnamefont
  {Fulvio}}, \bibinfo {author} {\bibfnamefont {I.}~\bibnamefont {Akinnola}}, \
  and\ \bibinfo {author} {\bibfnamefont {B.~R.}\ \bibnamefont {Postle}},\
  }\bibfield  {title} {\enquote {\bibinfo {title} {Gender (im)balance in
  citation practices in cognitive neuroscience},}\ }\href@noop {} {\bibfield
  {journal} {\bibinfo  {journal} {J Cogn Neurosci}\ }\textbf {\bibinfo {volume}
  {33}},\ \bibinfo {pages} {3--7} (\bibinfo {year} {2021})}\BibitemShut
  {NoStop}%
\bibitem [{\citenamefont {Zhou}\ \emph {et~al.}(2020)\citenamefont {Zhou},
  \citenamefont {Cornblath}, \citenamefont {Stiso}, \citenamefont {Teich},
  \citenamefont {Dworkin}, \citenamefont {Blevins},\ and\ \citenamefont
  {Bassett}}]{zhou_dale_2020_3672110}%
  \BibitemOpen
  \bibfield  {author} {\bibinfo {author} {\bibfnamefont {D.}~\bibnamefont
  {Zhou}}, \bibinfo {author} {\bibfnamefont {E.~J.}\ \bibnamefont {Cornblath}},
  \bibinfo {author} {\bibfnamefont {J.}~\bibnamefont {Stiso}}, \bibinfo
  {author} {\bibfnamefont {E.~G.}\ \bibnamefont {Teich}}, \bibinfo {author}
  {\bibfnamefont {J.~D.}\ \bibnamefont {Dworkin}}, \bibinfo {author}
  {\bibfnamefont {A.~S.}\ \bibnamefont {Blevins}}, \ and\ \bibinfo {author}
  {\bibfnamefont {D.~S.}\ \bibnamefont {Bassett}},\ }\href@noop {} {\enquote
  {\bibinfo {title} {Gender diversity statement and code notebook v1.0},}\ }
  (\bibinfo {year} {2020})\BibitemShut {NoStop}%
\bibitem [{\citenamefont {Ambekar}\ \emph {et~al.}(2009)\citenamefont
  {Ambekar}, \citenamefont {Ward}, \citenamefont {Mohammed}, \citenamefont
  {Male},\ and\ \citenamefont {Skiena}}]{ambekar2009name}%
  \BibitemOpen
  \bibfield  {author} {\bibinfo {author} {\bibfnamefont {A.}~\bibnamefont
  {Ambekar}}, \bibinfo {author} {\bibfnamefont {C.}~\bibnamefont {Ward}},
  \bibinfo {author} {\bibfnamefont {J.}~\bibnamefont {Mohammed}}, \bibinfo
  {author} {\bibfnamefont {S.}~\bibnamefont {Male}}, \ and\ \bibinfo {author}
  {\bibfnamefont {S.}~\bibnamefont {Skiena}},\ }\bibfield  {title} {\enquote
  {\bibinfo {title} {Name-ethnicity classification from open sources},}\ }in\
  \href@noop {} {\emph {\bibinfo {booktitle} {Proceedings of the 15th ACM
  SIGKDD international conference on Knowledge Discovery and Data Mining}}}\
  (\bibinfo {year} {2009})\ pp.\ \bibinfo {pages} {49--58}\BibitemShut
  {NoStop}%
\bibitem [{\citenamefont {Sood}\ and\ \citenamefont
  {Laohaprapanon}(2018)}]{sood2018predicting}%
  \BibitemOpen
  \bibfield  {author} {\bibinfo {author} {\bibfnamefont {G.}~\bibnamefont
  {Sood}}\ and\ \bibinfo {author} {\bibfnamefont {S.}~\bibnamefont
  {Laohaprapanon}},\ }\bibfield  {title} {\enquote {\bibinfo {title}
  {Predicting race and ethnicity from the sequence of characters in a name},}\
  }\href@noop {} {\bibfield  {journal} {\bibinfo  {journal} {arXiv preprint
  arXiv:1805.02109}\ } (\bibinfo {year} {2018})}\BibitemShut {NoStop}%
\bibitem [{\citenamefont {Rombach}\ \emph {et~al.}(2014)\citenamefont
  {Rombach}, \citenamefont {Porter}, \citenamefont {Fowler},\ and\
  \citenamefont {Mucha}}]{rombach2014core}%
  \BibitemOpen
  \bibfield  {author} {\bibinfo {author} {\bibfnamefont {M.~P.}\ \bibnamefont
  {Rombach}}, \bibinfo {author} {\bibfnamefont {M.~A.}\ \bibnamefont {Porter}},
  \bibinfo {author} {\bibfnamefont {J.~H.}\ \bibnamefont {Fowler}}, \ and\
  \bibinfo {author} {\bibfnamefont {P.~J.}\ \bibnamefont {Mucha}},\ }\bibfield
  {title} {\enquote {\bibinfo {title} {Core-periphery structure in networks},}\
  }\href@noop {} {\bibfield  {journal} {\bibinfo  {journal} {SIAM Journal on
  Applied mathematics}\ }\textbf {\bibinfo {volume} {74}},\ \bibinfo {pages}
  {167--190} (\bibinfo {year} {2014})}\BibitemShut {NoStop}%
\bibitem [{\citenamefont {{S. P. Borgatti, M. G.
  Everett}}(2000)}]{borgatti2000}%
  \BibitemOpen
  \bibfield  {author} {\bibinfo {author} {\bibnamefont {{S. P. Borgatti, M. G.
  Everett}}},\ }\bibfield  {title} {\enquote {\bibinfo {title} {{Models of
  core/periphery structures}},}\ }\href@noop {} {\bibfield  {journal} {\bibinfo
   {journal} {Social Networks}\ }\textbf {\bibinfo {volume} {21}},\ \bibinfo
  {pages} {375--395} (\bibinfo {year} {2000})}\BibitemShut {NoStop}%
\end{thebibliography}%
\end{document}